\documentclass[12pt,a4paper]{article}
\usepackage[pdftex]{graphicx}
\usepackage{bm}
\usepackage{amsmath, amssymb}
\usepackage{subfigure}
\usepackage{float}
\usepackage{xcolor}
\usepackage{natbib}
\usepackage{setspace}
\usepackage[bookmarksnumbered]{hyperref}

\def\blue{\color{blue}}

\begin{document}

\title{\bf An ECM algorithm for Skewed Multivariate Variance Gamma Distribution in Normal Mean-Variance Representation}
\author{Thanakorn Nitithumbundit\footnote{Corresponding author. Email: T.Nitithumbundit@maths.usyd.edu.au} \ and
    Jennifer S.K. Chan \\
    \small{School of Mathematics and Statistics, University of Sydney, NSW 2006, Australia}}
\date{\today}

\maketitle

{\bf Abstract:}
    Normal mean-variance mixture distributions are widely applied to simplify a model's implementation and improve their computational efficiency under the Maximum Likelihood (ML) approach.
    Especially for distributions with
    normal mean-variance mixtures representation such as the multivariate skewed variance gamma (MSVG) distribution, it utilises the expectation-conditional-maximisation (ECM) algorithm to iteratively obtain the ML estimates.
    To facilitate application to financial time series, the mean is further extended to include autoregressive terms.
    Techniques are proposed to deal with the unbounded density for small shape parameter and to speed up the convergence. Simulation studies are conducted to demonstrate the applicability of this model and examine estimation properties.
    Finally, the MSVG model is applied to analyse the returns of five daily closing price market indices and standard errors for the estimated parameters are computed using Louis's method. \par \vspace{2mm}


\section{Introduction}

Variance Gamma (VG) distribution has been widely used to model financial time series data, particularly the log-price increment (return) which has the characteristics of having high concentration of data points around the center with occasional extreme values.
Some key properties includes finite moments of all orders and a member of the family of elliptical distributions in the symmetric case. See \cite{MadanSeneta1990} for more details of the properties of the VG distribution and its financial applications.
However, its applicability is still limited by its rather complicated density.
To estimate the model parameters \cite{MadanSeneta1987cheby} applied the ML approach using characteristic function from a Chebyshev polynomial expansion.
\cite{MadanSeneta1990} and \cite{Tjetjep2006} used the method of moments to estimate the parameters for the VG and skewed VG models respectively.
\cite{Finlay2008} compared the performance of various estimation methods for data with long range of dependence, namely the methods of moments, product-density maximum likelihood, minimum $\chi^2$, and the empirical characteristic function.
They showed that the performance of product-density maximum likelihood estimator is better than the method of moments and minimum $\chi^2$ estimators in terms of goodness-of-fit.
\cite{McNeil2005} (\S 3.2.4) applied the Expectation-Maximisation (EM) algorithm to generalised hyperbolic (GH) distribution which nests VG distribution as a special case.
However, they did not address the main challenge in estimation when the density is unbounded at the center, a feature of some financial time series. \cite{Fung2010} adopted a Bayesian approach with diffuse priors as a proxy for ML approach to estimate the bivariate skewed VG and Student-t models.
They implemented the models using the Bayesian software \texttt{WinBUGS}, and compared their goodness-of-fit performances with simulated and real data sets.
We propose to use EM algorithm under the ML approach to estimate parameters
of the VG model directly. Our proposed method not only provides improved accuracy but also handles the problem of unbounded density. \par \vspace{3mm}

It is well known that the VG distribution has normal mean-variance mixtures representation in which the data follow a normal distribution condition on some mixing parameters which have a gamma distribution.
Large values of the mixing parameters correspond to the normal distributions having relatively larger variances to accommodate outliers.
Hence outlier diagnosis can be performed using these mixing parameters \citep{Choy1997Hierarchical}.
More importantly, the conditional normal data distribution greatly simplifies the parameter estimation based on standard results.
However, the mixing parameters are not observed.
In case with missing data, ML estimation becomes very challenging as the marginal likelihood function involves high dimensional integration.
\cite{Dempster1977} showed that the EM algorithm can be used
    to find the ML estimates for univariate Student-t distribution with fixed degrees of freedom in normal mean-variance mixture representation.
    Dempster, Laird and Rubin (1980) extended these results to the regression case.
\cite{Meng1993} improved the EM algorithm to Expectation/Conditional Maximisation (ECM)
    algorithm which simplifies the maximisation step by utilising some standard results of normal distribution when conditioned on some mixing parameters.
    Moreover, the ECM algorithm still possesses desirable properties from the EM algorithm.
\cite{Meng1994}
    considered a
     variation of the ECM algorithm called multicycle ECM (MCECM) which inserts extra E-steps after each update of parameters.
\cite{Liu1994}
    advanced the ECM algorithm to Expectation/Conditional Maximisation Either (ECME) by maximising the actual likelihood rather than the conditional/constrained expected likelihood.
\cite{Liu1995}
    applied the MCECM and ECME algorithms to obtain the ML estimates for multivariate Student-t distribution with incomplete data. They also found that the ECME algorithm converges much more efficiently than the EM and ECM algorithms in terms of computational time.
\cite{Hu2008}
    used the MCECM algorithm with the Student-t distribution for portfolio credit risk measurement.
\par \vspace{3mm}

We adopt the ECM algorithms for VG distribution and extend it to multivariate skewed VG (MSVG) distribution because univariate symmetric VG distribution fails to account for the dynamics of cross-correlated time series with asymmetric tails.
For example, the returns of daily price for stocks in related industries are cross-correlated across stocks, serially correlated over time and possibly right or left tailed. We further include autoregressive (AR) terms in the means to allow for autocorrelation.
This generalised model called MSVG AR can describe many essential features of financial time series.
However this model, as well as its estimation methodologies that can cope with various computational difficulties in ML implementation, is still relatively scarce in the literature.
For the ECM algorithms, these computational difficulties include accurately calculating the ML estimates with unbounded likelihood while maintaining computational efficiency.
\par \vspace{3mm}
To deal with these technical issues and evaluate the performance ECM algorithms, we perform three simulation studies.
Firstly, as the two ECM type algorithms, namely the MCECM and ECME algorithms, have a trade-off between computational efficiency and performance, we derive a method we call the hybrid ECM (HECM) to improve the efficiency whilst also maintaining good performance.
We compare the efficiency of these three algorithms through simulated data.
Secondly, we propose to bound the density within a certain range around the centre should it become unbounded when the shape parameter is small and observations are close to the mean.
We show that the proposed technique improves computational efficiency and performance of the HECM algorithm.
Lastly, we analyse the performance of the algorithm for different MSVG distributions with high or low levels of shape and skew parameters.
Results are promising. We then apply the HECM estimator to fit the MSVG model to multiple stock market indices.
To assess the significance of  parameter estimates, standard errors are also evaluated using
Louis's method \citep{Louis1982}.
Results show that the MSVG AR model provides good fit to the data and reveals high correlation between some pairs of indices.
We then fit a bivariate model to a pair of  highly correlated market indices.
The contour plot of fitted density reveals that the bivariate model captures the high concentration of observations in the center and heavy outliers on the edge. Results facilitate portfolio setting based on a basket of market indices. \par \vspace{3mm}

In summary, our proposed HECM estimator is new in the literature of VG models and forms a useful toolkit to its implementation with all important technical issues clearly addressed. We also showcase its efficiency and demonstrate its applicability through real examples. Nevertheless the techniques provided in HECM estimation including the way to handle unbounded density can be generalised to other distributions with normal mean-variance mixtures representation.
The structure of the paper is as follows.
Section \ref{Section: msVG model} introduces the MSVG distribution and some of its key properties.
Section \ref{Section: ECM algorithm} presents the ECM algorithms for estimating the model parameters and the calculation of their standard errors by computing the information matrix through Louis's formula.
We also propose some techniques to address the technical issues in the ECM algorithms.
Section \ref{Section: Simulation study} describes the simulation studies to assess the performance of proposed techniques in handling these technical issues.
Section \ref{Section: Real data analysis} implements the HECM algorithm to real data.
Finally, a brief conclusion with discussion is given in Section \ref{Section: Conclusion/Discussion}.

\section{Multivariate Skewed Variance Gamma Model}      \label{Section: msVG model}
While univariate VG models has been considered in many times in the literature, MSVG models are in fact more applicable because
they reveal the dependence between different components and allows asymmetric tail behavior.
Let $\bm y_i, \ i=1, \dots, n$
be a set of $d$-dimensional multivariate observations following a MSVG distribution with the probability density function (pdf) given by: \small
\begin{align} \label{VGpdf}
    f_{VG} (\bm y)
        =& \ \frac{2^{1-\nu} \nu^{\frac{d}{2}}}
                    {\left|\bm\Sigma\right|^{\frac{1}{2}} \pi^{\frac{d}{2}} \Gamma\left(\nu\right)}
                \frac{K_{\nu-\frac{d}{2}}\left(\sqrt{[2\nu + \bm \gamma'\bm\Sigma^{-1}\bm \gamma]
                (\bm y - \bm\mu)'\bm\Sigma^{-1}(\bm y - \bm\mu) }\right)
                \exp\left((\bm y - \bm\mu)'\bm\Sigma^{-1}\bm \gamma\right)}
                    {[(2\nu + \bm \gamma'\bm\Sigma^{-1}\bm \gamma )(\bm y - \bm\mu)'\bm\Sigma^{-1}(\bm y - \bm\mu)]^{-\frac{2\nu-d}{4}}
                        [1 + \frac{1}{2\nu}\bm \gamma'\bm\Sigma^{-1}\bm \gamma]^\frac{2\nu-d}{2}}
\end{align}
\normalsize
where $\bm\mu\in\mathbb R^d$ is the location parameter,
$\bm\Sigma$ is a $d\times d$ positive definite symmetric scale matrix,
$\bm\gamma\in\mathbb R^d$ is the skewness parameter, $\nu>0$ is the shape parameter, $\Gamma(\cdot)$ is the gamma function and
$K_\eta(\cdot)$ is the modified Bessel function of the second kind with index $\eta$
\citep[see][\S9.6]{Abramowitz2007}. 
 Based on this parametrisation, decreasing the shape parameter $\nu$ will increase the probability around $\bm \mu$ as well as the tail probabilities at the expense of probability in the intermediate range. See  \cite{Fung2007} for more information about the shape parameter $\nu$ and the tail behavior which is of power-modified exponential-type.
\par \vspace{3mm}

The MSVG distribution has a normal mean-variance mixtures representation given by:
\begin{equation}
    \bm y_i | \lambda_i \sim \mathcal N_d(\bm\mu+\bm\gamma\lambda_i, \lambda_i \bm\Sigma), \quad \lambda_i \sim \mathcal G(\nu,\nu) \label{scale mixture rep of VG}
\end{equation}
where $\mathcal G(\alpha,\beta)$ is a Gamma distribution with shape parameters $\alpha>0$, rate parameter $\beta>0$ and pdf:
\begin{equation*}
    f_G(\lambda) = \frac{\beta^\alpha}{\Gamma(\alpha)} \lambda^{\alpha-1} \exp(-\beta \lambda) , \text{ for } \lambda>0.
\end{equation*}
The mean and variance of a MSVG random vector $\bm Y_i$ are given by:
\begin{align}
    \mathbb E (\bm Y_i) = \bm\mu + \bm\gamma \quad \text{and} \quad
    \mathbb C\text{ov}(\bm Y_i) = \bm\Sigma + \tfrac{1}{\nu} \bm\gamma \bm\gamma',    \label{CovY}
\end{align}
respectively. , The pdf in \eqref{VGpdf} as $\bm y_i \rightarrow \bm\mu$ is given by:
\begin{align*}
    f_{VG}(\bm y_i) \sim
        \begin{cases}
            \displaystyle2^{\nu-d} \pi^{-\frac{d}{2}} \left|\bm\Sigma\right|^{-\frac{1}{2}}
                \frac{\Gamma \left( \nu - \tfrac{d}{2} \right)}{\Gamma\left(\nu\right)}
                \frac{ \nu^\nu  }
                    {(2\nu + \bm \gamma'\bm\Sigma^{-1}\bm \gamma)^\frac{2\nu-d}{2}}
                & \text{ if } \nu>\tfrac{d}{2}, \\ \vspace{2mm}
            \displaystyle-2^{1-d} \pi^{-\frac{d}{2}} \left|\bm\Sigma\right|^{-\frac{1}{2}}
                \frac{d^\frac{d}{2}}{\Gamma\left(\tfrac{d}{2}\right)}
                \log \left(\delta_i \right)
                & \text{ if } \nu=\tfrac{d}{2}, \\ \vspace{2mm}
            \displaystyle2^{-\nu} \pi^{-\frac{d}{2}} \left|\bm\Sigma\right|^{-\frac{1}{2}}
                \frac{\Gamma \left(\tfrac{d}{2} - \nu \right)}{\Gamma\left(\nu\right)}
                \frac{\nu^\nu}{\delta_i^{d-2\nu}}
                & \text{ if } \nu<\tfrac{d}{2},
        \end{cases}
\end{align*}
where $\delta^2_i = (\bm y_i - \bm\mu)'\bm\Sigma^{-1}(\bm y_i - \bm\mu)$. This shows that the pdf at $\bm\mu$ is unbounded when $\nu\leq \frac{d}{2}$. This is an important property of the MSVG distribution to be addressed in the next section.

\begin{figure}[htbp]
  \begin{center}
    \subfigure[standard contour plot]{\label{}\includegraphics[scale=0.55]{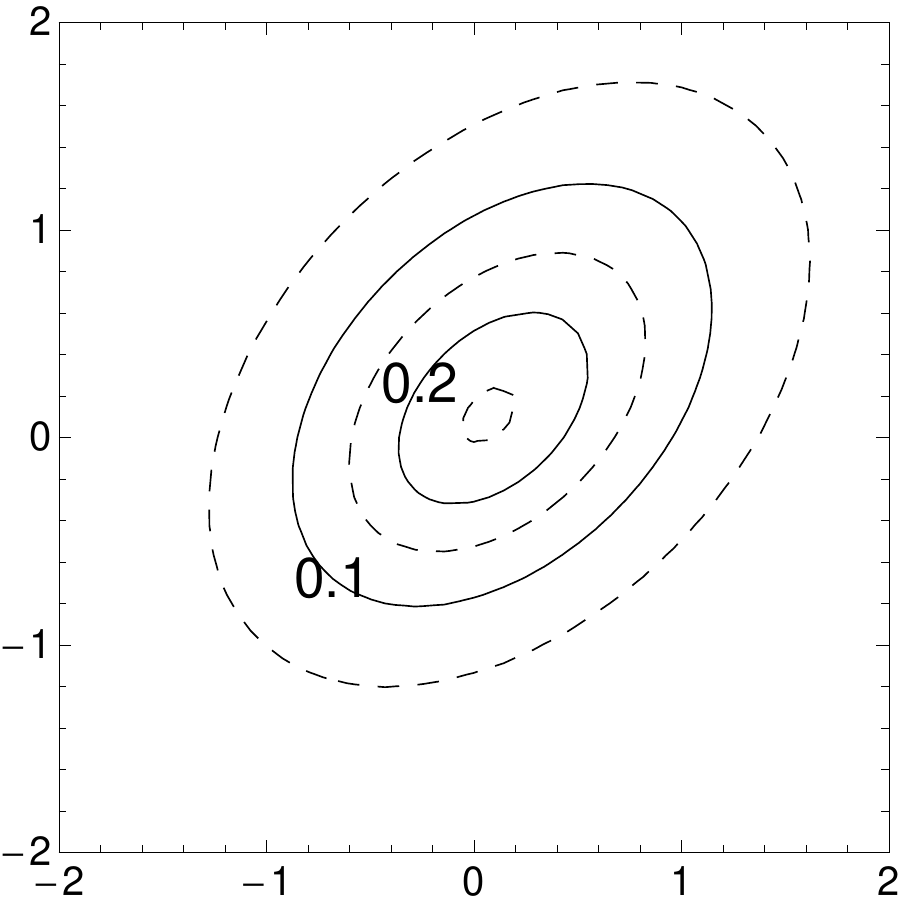}}
    \subfigure[small $\nu$ contour plot]{\label{}\includegraphics[scale=0.55]{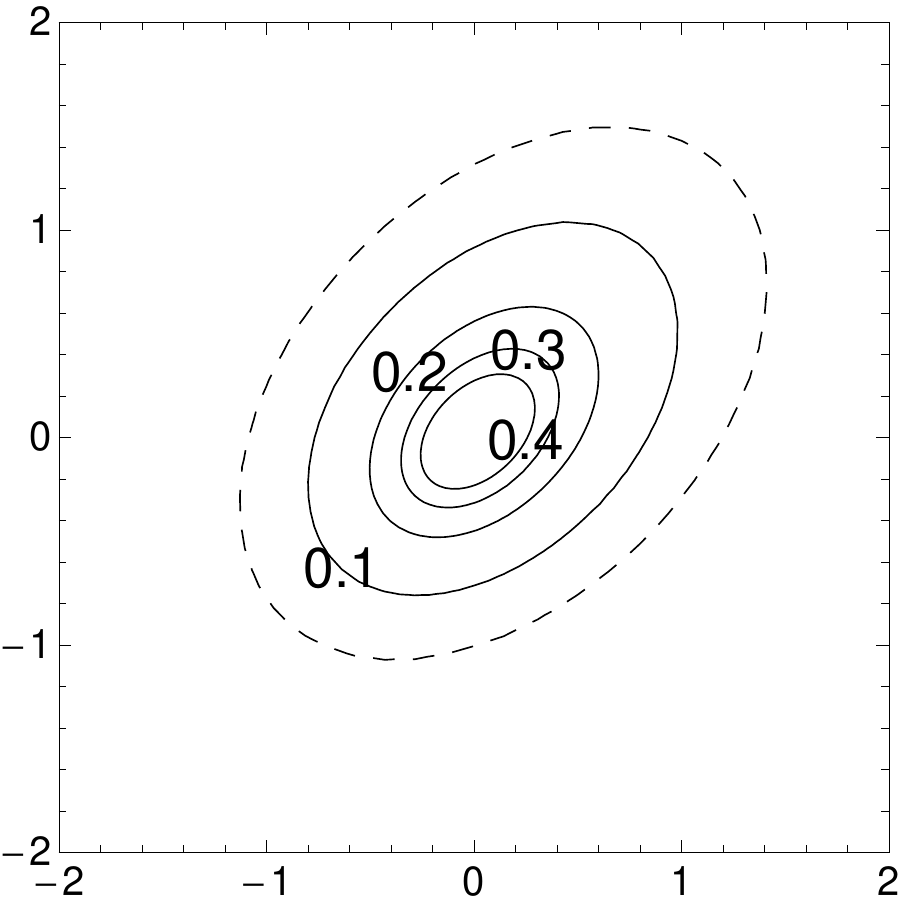}} \\
    \subfigure[standard 3D plot]{\label{}\includegraphics[scale=0.16]{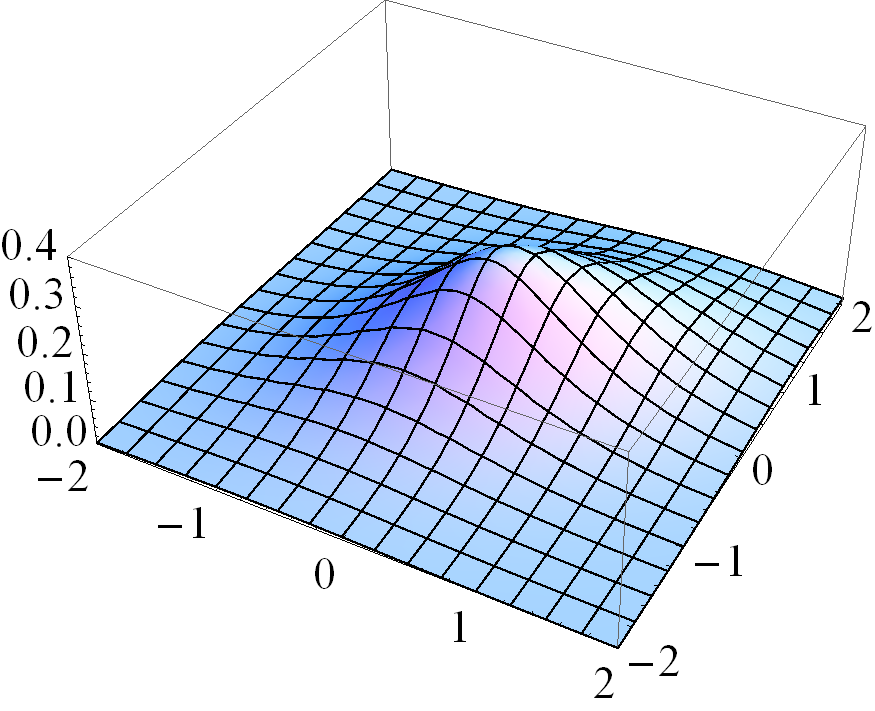}}
    \subfigure[small $\nu$ 3D plot]{\label{}\includegraphics[scale=0.16]{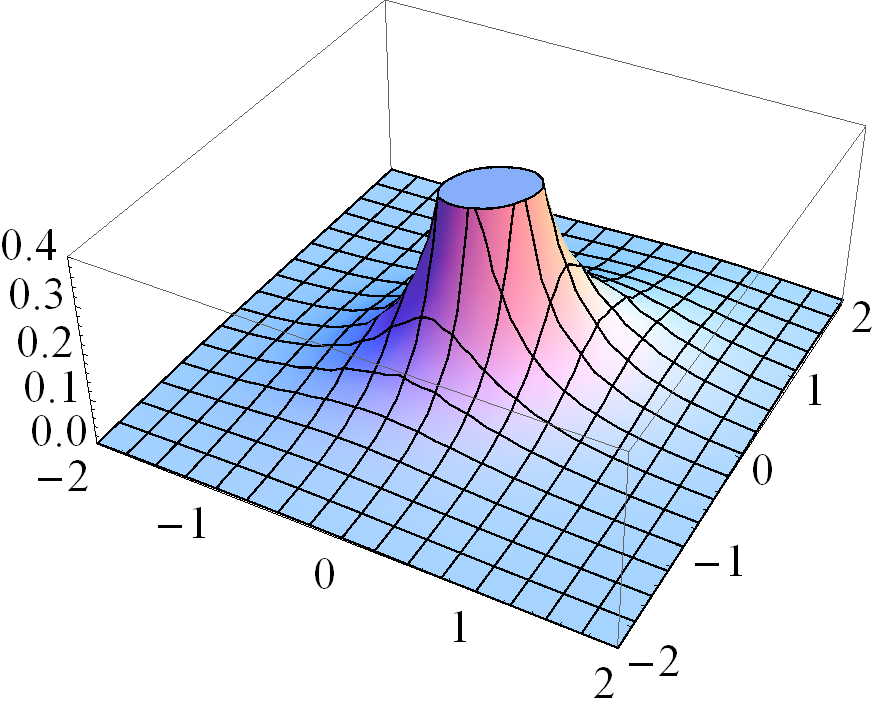}} \\
    \subfigure[large $\bm\gamma$ contour plot]{\label{}\includegraphics[scale=0.55]{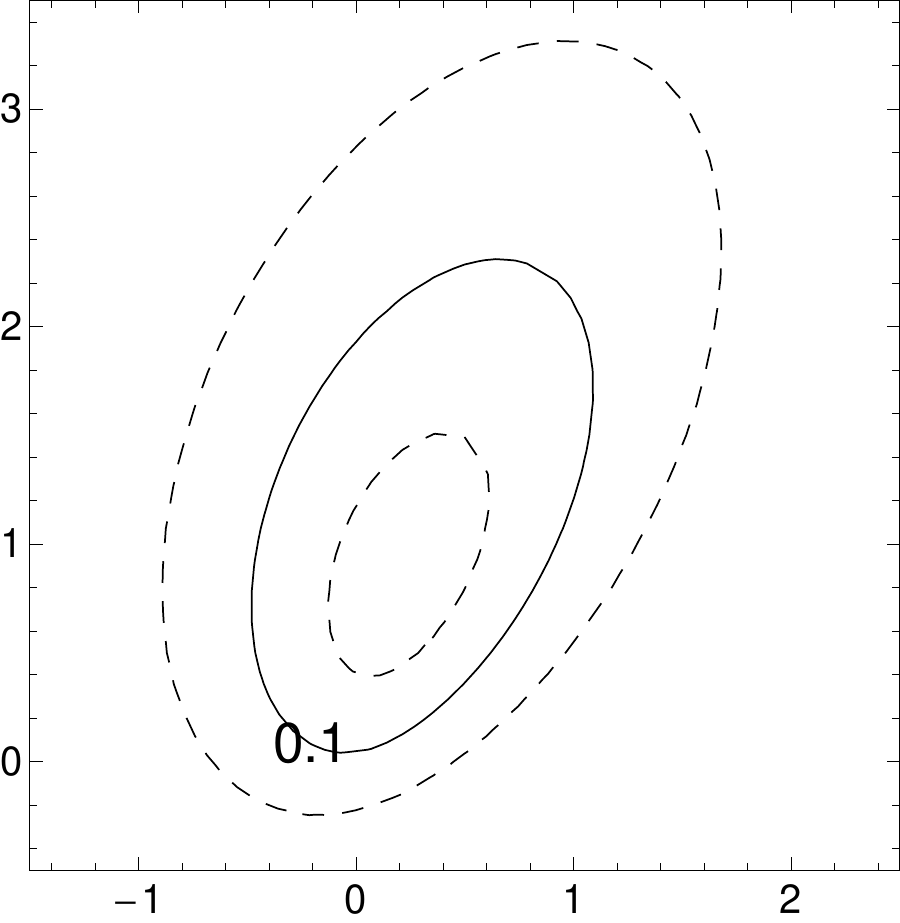}}
    \subfigure[large correlation contour plot]{\label{}\includegraphics[scale=0.55]{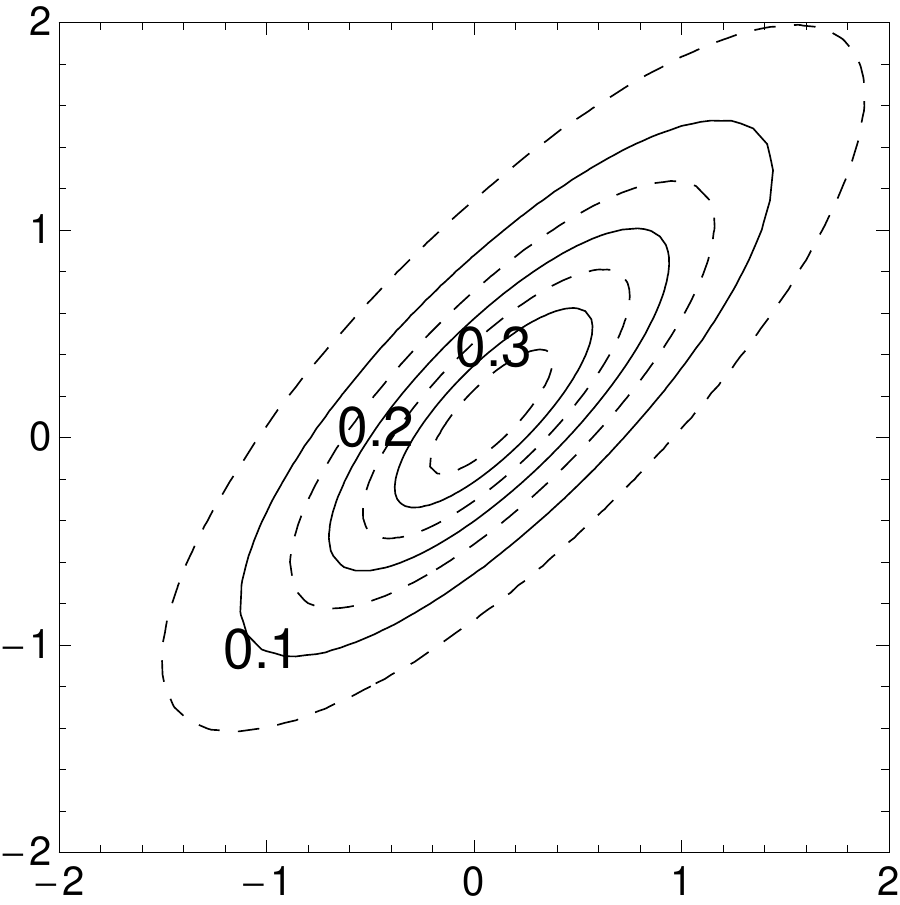}} \\
    \subfigure[large $\bm\gamma$ 3D plot]{\label{}\includegraphics[scale=0.16]{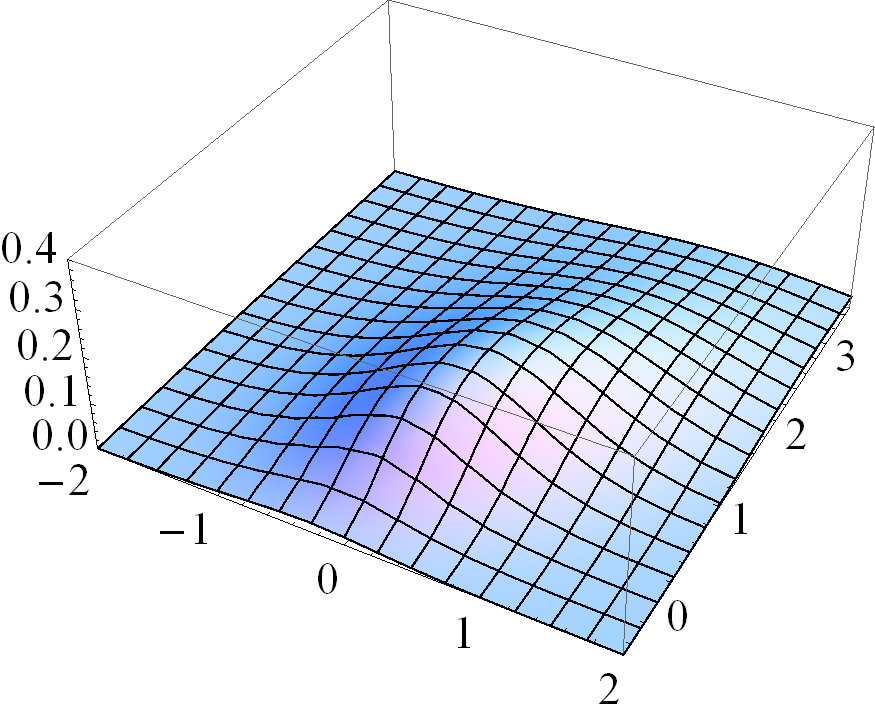}}
    \subfigure[large correlation 3D plot]{\label{}\includegraphics[scale=0.16]{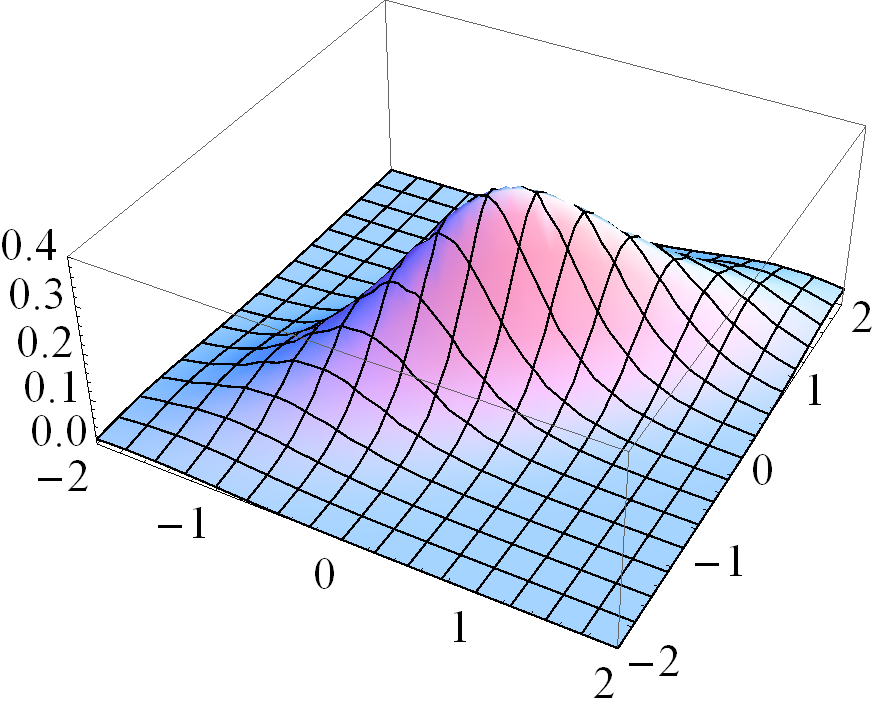}}
  \end{center}
  \caption{Various contour and 3D plots of bivariate skewed VG model with for different parameters.
        In the contour plots, the bold lines represent level sets $\{0.1,0.2,0.3,0.4\}$, and
            the dashed lines represent level sets $\{0.05,0.15,0.25,0.35\}$.
        The range for the 3D plots is kept between $0$ and $0.4$}
  \label{msVGplots}
\end{figure}

Figure \ref{msVGplots} gives four pairs of contour and three dimensional plots for various cases of the MSVG distribution.
The first pair of plots is based on parameters
\begin{equation*}
\bm\mu=
    \begin{pmatrix}
        0 \\
        0
    \end{pmatrix} , \quad
\bm\Sigma=
    \begin{pmatrix}
        1 & 0.4 \\
        0.4 & 1
    \end{pmatrix} , \quad
\bm\gamma=
    \begin{pmatrix}
        0.2 \\
        0.3
    \end{pmatrix} , \ \text{and} \ \quad
\nu=3.
\end{equation*}
Based on the distribution for the first pair of plots, three other pairs of plots demonstrate the changes in pdf when the shape parameter decreases to $\nu=0.6$, the skewness parameter increases to $\bm\gamma=(0.5,2)$, and the correlation coefficient in $\bm \Sigma$ increases to 0.8, respectively,  while keeping other parameters fixed.
Note that the density for the case with small $\nu$ is unbounded.

\section{ECM Algorithm}     \label{Section: ECM algorithm}
In the likelihood approach, maximising the complicated density in \eqref{VGpdf} can be computationally intensive.
We propose the ECM algorithm in the likelihood approach because the normal mean-variance mixtures representation in \eqref{scale mixture rep of VG} can greatly simplify the maximisation procedures in the M-step of the algorithm.
Then we extend the algorithm to incorporate an AR(1) term and discuss some technical issues involving small $\nu$. \par \vspace{3mm}

\subsection{Basic algorithm} \label{Subsection: Basic algorithm}
The ML estimates for parameters $\bm\theta=(\bm\mu,\bm\Sigma,\bm\gamma,\nu)$ in the parameter space $\bm\Theta$ maximises the log-likelihood
\begin{equation*}
    \ell(\bm\theta| \bm y_1,...,\bm y_n) = \sum_{i=1}^n \log f_{VG}(\bm y_i| \bm\theta)
\end{equation*}
where $f_{VG}(\cdot)$ is the pdf of MSVG distribution in \eqref{VGpdf}. Using the normal mean-variance mixtures representation of the MSVG distribution in \eqref{scale mixture rep of VG}, we consider
$\bm\lambda = \{ \lambda_1,...,\lambda_n\}$ to be unobserved and $\{\bm y,\bm \lambda\}$ to be the complete data where $\bm y = (\bm y_1, ..., \bm y_n)$.
The complete data density can be represented as a product of conditional normal densities given the unobserved mixing parameters and the gamma densities of the mixing parameters, that is,
\begin{equation*}
    f(\bm y, \bm\lambda|\bm \theta)
        = \prod_{i=1}^n f_N(\bm y_i | \lambda_i,\bm\mu,\bm\Sigma,\bm\gamma) f_G(\lambda_i|\nu).
\end{equation*}
This is essentially a state space model with normals as the structural distributions, $\lambda_i$ as the state or missing parameters, and gamma as the mixing distribution. Due to the mixing structure, the complete-data log-likelihood function can be factorised into two distinct functions as follows:
\begin{align} \label{decomposed loglik}
    \ell(\bm\theta | \bm y , \bm\lambda)
        = \ell_N(\bm\mu, \bm\Sigma, \bm\gamma|\bm y , \bm\lambda)
            + \ell_G(\nu|\bm\lambda)
\end{align}
where the log-likelihood of the conditional normal distributions is given by:
\begin{align}  \label{Npdf}
    \ell_N(\bm\mu, \bm\Sigma, \bm\gamma|\bm y , \bm\lambda)
        = -\frac{n}{2}\log|\bm\Sigma| - \frac{1}{2}\sum_{i=1}^n
                \frac{1}{\lambda_i}(\bm y_i - \bm\mu - \lambda_i\bm\gamma)'
                    \bm\Sigma^{-1} (\bm y_i - \bm\mu - \lambda_i\bm\gamma)+C_1
\end{align}
for some constant $C_1$ and the log-likelihood of the conditional gamma distributions is given by:
\begin{align}
    \ell_G(\nu|\bm\lambda)  \label{Gamma log-likelihood}
        = n\nu\log\nu - n\log\Gamma(\nu) + (\nu-1)\sum_{i=1}^n \log\lambda_i - \nu\sum_{i=1}^n \lambda_i.
\end{align}
Hence, condition on $\bm \lambda$, the estimation of all parameters $(\bm\mu,\bm\Sigma,\bm\gamma,\nu)$ can be separated in two blocks:
the conditional maximisation (CM) of $(\bm\mu,\bm\Sigma,\bm\gamma)$ from the conditional normals log-likelihood function and the CM of $\nu$ from the gamma log-likelihood function.
The mixing parameter $\bm \lambda$ is first estimated by the conditional expectation given the observed-data $\bm y$. The procedures are described as below: \par \vspace{3mm}

\noindent {\bf E-step for $\lambda_i$:} \\
To derive the conditional expectations of the
$\lambda_i$'s given the observed-data $\bm y_i$ and parameters $(\bm\mu,\bm\Sigma,\bm\gamma,\nu)$,
we need the conditional posterior distribution of $\lambda_i$ given by:
\begin{align} \label{pdfGIG}
f(\lambda_i| \bm y_i, \bm\mu, \bm\Sigma, \bm\gamma,\nu)
    \propto& \ f(\lambda_i, \bm y_i| \bm\mu, \bm\Sigma, \bm\gamma, \nu) \nonumber \\
    \propto& \ \lambda_i^{\nu-\frac{d}{2}-1}
        \exp\left[-\frac{1}{2\lambda_i}(\bm y_i - \bm\mu)'\bm\Sigma^{-1}(\bm y_i - \bm\mu)
            - \frac{\lambda_i}{2}\left(\bm\gamma'\bm\Sigma^{-1}\bm\gamma + 2\nu\right)\right]
\end{align}
which corresponds to the pdf of a generalised inverse Gaussian distribution \citep{Embrechts1983}. Using this posterior distribution, we can calculate the following conditional expectations:
\begin{align}
    \widehat{\lambda_i} \label{E-step lambda1}
        =& \ \mathbb{E} \left(\lambda_i | \bm y, \bm\theta \right)
        =  \frac{\delta_i K_{\nu-\frac{d}{2}+1}\left(\sqrt{2\nu + \bm\gamma'\bm\Sigma^{-1}\bm\gamma} \delta_i\right)}
                {\sqrt{2\nu+\bm\gamma'\bm\Sigma^{-1}\bm\gamma} K_{\nu-\frac{d}{2}}\left(\sqrt{2\nu+\bm\gamma'\bm\Sigma^{-1}\bm\gamma}\delta_i\right)} , \\
    \widehat{1/\lambda_i} \label{E-step lambda2}
        =&\ \mathbb{E} \left(\frac{1}{\lambda_i} \Bigg| \bm y, \bm\theta \right)
        =  \frac{\sqrt{2\nu+\bm\gamma'\bm\Sigma^{-1}\bm\gamma} K_{\nu-\frac{d}{2}-1}\left(\sqrt{2\nu+\bm\gamma'\bm\Sigma^{-1}\bm\gamma}\delta_i\right)}
                {\delta_i K_{\nu-\frac{d}{2}}\left(\sqrt{2\nu+\bm\gamma'\bm\Sigma^{-1}\bm\gamma}\delta_i\right)} , \\
    \widehat{\log\lambda_i} \label{E-step lambda3}
        =& \ \mathbb{E} \left(\log\lambda_i | \bm y, \bm\theta \right)
        = \log\left(\frac{\delta_i}{\sqrt{2\nu+\bm\gamma'\bm\Sigma^{-1}\bm\gamma}}\right)
            + \frac{K_{\nu-\frac{d}{2}}^{(1,0)}(\sqrt{2\nu+\bm\gamma'\bm\Sigma^{-1}\bm\gamma}\delta_i)}
                {K_{\nu-\frac{d}{2}}(\sqrt{2\nu+\bm\gamma'\bm\Sigma^{-1}\bm\gamma}\delta_i)}
\end{align}
where
$\displaystyle K_{\nu}^{(1,0)}(z) = \tfrac{\partial}{\partial\alpha} K_\alpha (z)\big|_{\alpha=\nu}$
which is approximated using the central difference formula
\begin{equation}    \label{central difference formula for BesselK}
    K_{\nu}^{(1,0)}(z) \approx \frac{K_{\nu+h}(z) - K_{\nu-h}(z)}{2h}
\end{equation}
where we let $h=10^{-5}$.
\par \vspace{6mm}

\noindent {\bf CM-step for $(\bm\mu,\bm\Sigma,\bm\gamma)$:} \\
Given the mixing parameters $\bm\lambda$, the ML estimate of $(\bm\mu, \bm\Sigma, \bm\gamma)$ can be obtained by maximising $\ell_N(\bm\mu, \bm\Sigma, \bm\gamma|\bm y , \bm\lambda)$ in \eqref{Npdf} with respect to $(\bm\mu, \bm\Sigma, \bm\gamma)$.
After equating each component of the partial derivatives of $\ell_N(\bm\mu, \bm\Sigma, \bm\gamma|\nu, \bm y , \bm\lambda)$ to zero, we obtain the following estimates:
\begin{align}
    \hat{\bm\mu} \label{CM-step mu}
        =& \frac{S_{\bm y/\lambda} S_{\lambda} - n S_{\bm y}}
                {S_{1/\lambda} S_{\lambda} - n^2} , \\
    \hat{\bm\gamma} \label{CM-step gamma}
        =& \frac{S_{\bm y} - n\hat{\bm\mu}}{S_{\lambda}} , \\
    \hat{\bm\Sigma} \label{CM-step Sigma}
        =& \frac{1}{n} \sum_{i=1}^n \frac{1}{\lambda_i} (\bm y_i - \hat{\bm\mu})(\bm y_i - \hat{\bm\mu})'
            - \frac{1}{n} \hat{\bm\gamma}\hat{\bm\gamma}' S_{\lambda},
\end{align}
where the complete data sufficient statistics are
\begin{align} \label{suffstats}
    S_{\bm y} = \sum_{i=1}^n \bm y_i , \quad
    S_{\bm y/\lambda} = \sum_{i=1}^n \frac{1}{\lambda_i} \bm y_i , \quad
    S_{\lambda} = \sum_{i=1}^n \lambda_i , \quad
    S_{1/\lambda}  = \sum_{i=1}^n \frac{1}{\lambda_i}.
\end{align} \par \vspace{3mm}

\noindent {\bf CM-step for $\nu$:} \\
Given the mixing parameters $\bm\lambda$,
the ML estimate of $\nu$ can be obtained by numerically maximising $\ell_G(\nu|\bm\lambda)$ in \eqref{Gamma log-likelihood} with respect to $\nu$
using Newton-Raphson (NR) algorithm where the derivatives is given by:
\begin{align}
    \label{NR1} \frac{\partial\ell}{\partial\nu} = & \ n + n\log\nu -n\psi(\nu) + S_{\log \lambda} - S_\lambda,\\
    \label{NR2} \frac{\partial^2\ell}{\partial\nu^2} = & \ \frac{n}{\nu} - n\psi'(\nu).
\end{align}
where $\psi(x)=\frac{d}{dx} \log \Gamma(x)$ is the digamma function and
$S_{\log\lambda} = \sum_{i=1}^n \log\lambda_i$.
This ECM algorithm is the MCECM algorithm. In summary, the algorithm involves the following steps: \par \vspace{2mm} \noindent
{\bf Initialisation step:} Choose suitable starting values $(\bm\mu_0, \bm\Sigma_0, \bm\gamma_0, \nu_0)$ for the parameters. If the data is roughly symmetric around the mean, then it is recommended to choose starting values
$(\bar {\bm y}, \bm Q_{y},\bm 0,d)$
where $\bar {\bm y}$ and $\bm Q_{y}$ denote the sample mean and sample variance-covariance matrix of $\bm y$ respectively. \par \vspace{3mm}

\noindent At the $t$-th
iteration with current estimates
$(\bm\mu^{(t)}, \bm\Sigma^{(t)}, \bm\gamma^{(t)}, \nu^{(t)})$, \par \vspace{3mm}

\noindent {\bf E-step 1:}
    Calculate $\widehat{\lambda}_i^{(t+1/2)}$ and $\widehat{1/\lambda_i}^{(t+1/2)}$ for $i=1,...,n$
    in \eqref{E-step lambda1} and \eqref{E-step lambda2}, respectively, using
            $(\bm\mu^{(t)}, \bm\Sigma^{(t)}, \bm\gamma^{(t)}, \nu^{(t)})$.
    Calculate also the sufficient statistics
        $S_{\bm y/\lambda}^{(t+1/2)}$,
        $S_{\lambda}^{(t+1/2)}$ and
        $S_{1/\lambda}^{(t+1/2)}$ in \eqref{suffstats}. \par \vspace{3mm}

\noindent {\bf CM-step 1:} Update the estimates to
        $(\bm\mu^{(t+1)}, \bm\Sigma^{(t+1)}, \bm\gamma^{(t+1)})$
    in \eqref{CM-step mu} to \eqref{CM-step Sigma} respectively using the sufficient statistics. \par \vspace{3mm}

\noindent {\bf E-step 2:}
    Calculate $\widehat{\lambda_i}^{(t+1)}$ and $\widehat{\log\lambda_i}^{(t+1)}$ for $i=1,...,n$
    in \eqref{E-step lambda1} and \eqref{E-step lambda3} respectively using the updated estimates
        $(\bm\mu^{(t+1)}, \bm\Sigma^{(t+1)}, \bm\gamma^{(t+1)}, \nu^{(t)})$.
    Calculate also the sufficient statistics
        $S_{\lambda}^{(t+1)}$ and $S_{\log\lambda}^{(t+1)}$
    in \eqref{suffstats}. \par \vspace{3mm}

\noindent {\bf CM-step 2:} Update the estimate to $\nu^{(t+1)}$
    using the derivatives in \eqref{NR1} and \eqref{NR2} for the NR procedure and the sufficient statistics. \par \vspace{3mm}

\noindent {\bf Stopping rule:} Repeat the procedures until the relative increment of log-likelihood function is smaller than the tolerance level $10^{-8}$. \par \vspace{3mm}

Alternatively, the ECME algorithm update the estimate to $\nu^{(t+1)}$ in CM-step 2 by maximising directly the actual log-likelihood
$\ell_{VG}(\nu | \bm y, \bm\mu, \bm\Sigma, \bm\gamma)$ which is the sum of log pdfs in \eqref{VGpdf} for $(\bm y_1, \dots, \bm y_n)$. This procedure is implemented using the \texttt{optimize} function specified in \texttt{R} which
is a combination of golden section search and successive parabolic interpolation.

\subsection{Algorithm with AR term}       \label{Subsection: Algorithm with AR(1) term}

Suppose now our observed-data is $\bm y = (\bm y_0, ..., \bm y_n)$.
The MSVG distribution for $\bm y_i$ where $i=1,...,n$ with AR mean function of order 1 (without loss of generosity) can be represented hierarchically as follows:
\begin{equation}
    \bm y_i | \lambda_i \sim \mathcal N_d(\bm\beta_0 + \bm\beta_1 \bm y_{i-1}+\bm\gamma\lambda_i, \lambda_i \bm\Sigma), \quad
    \lambda_i \sim \mathcal G(\nu,\nu) \label{scale mixture rep of VG with AR(1)}
\end{equation}
where $\bm\beta_0 \in \mathbb R^d$, and $\bm\beta_1$ is a $d\times d$ matrix with all eigenvalues less than 1 in modulus.
Because of the dependency structure, we maximise the conditional likelihood function for $\bm y_1, ..., \bm y_n$ given $\bm y_0$ expressed by:
\begin{align*}
    f(\bm y_1,...,\bm y_n,\lambda_1,...,\lambda_n|\bm y_0 ; \bm\theta)
        =& \prod_{i=1}^n f_N(\bm y_i|\lambda_i, \bm y_{i-1}; \bm\theta) f_G(\lambda_i;\nu).
\end{align*}
So the conditional log-likelihood is given by:
\begin{equation*}
    \ell_c(\bm\theta|\bm y, \bm\lambda)
        = \ell_N(\bm\beta_0,\bm\beta_1,\bm\Sigma,\bm\gamma|\bm y,\bm\lambda)
            + \ell_G(\nu|\bm\lambda)
\end{equation*}
where the log-likelihood of the conditional normal distribution is given by:
\begin{align}
    &\ell_N(\bm\beta_0,\bm\beta_1,\bm\Sigma,\bm\gamma|\bm y,\bm\lambda) \nonumber \\
        &\hspace{1cm} = -\frac{n}{2}\log|\bm\Sigma| - \frac{1}{2}\sum_{i=1}^n
                \frac{1}{\lambda_i}(\bm y_i - \bm\beta_0 - \bm\beta_1 \bm y_{i-1} - \lambda_i\bm\gamma)'
                    \bm\Sigma^{-1} (\bm y_i - \bm\beta_0 - \bm\beta_1 \bm y_{i-1} - \lambda_i\bm\gamma) + C_2 \label{loglikN with AR(1)}
\end{align}
for some constant $C_2$ and $\ell_G(\nu|\bm\lambda)$ is given in equation \eqref{Gamma log-likelihood}. \\

\noindent {\bf E-step for $\lambda_i$:} \\
This step is similar to the E-step in Section \ref{Subsection: Basic algorithm}. Just replace
$\delta_i^2 = \left(\bm y_i-\bm\mu\right)'\bm\Sigma^{-1}\left(\bm y_i-\bm\mu\right)$ with
$\delta_i^2 = \left(\bm y_i-\bm\beta_0-\bm\beta_1\bm y_{i-1}\right)'\bm\Sigma^{-1}\left(\bm y_i-\bm\beta_0-\bm\beta_1 \bm y_{i-1}\right)$. \\

\noindent {\bf CM-step for $(\bm\beta_0,\bm\beta_1,\bm\Sigma,\bm\gamma)$:} \\
Similarly condition on $\lambda_i$, the ML estimates of
$(\bm\beta_0, \bm\beta_1, \bm\Sigma, \bm\gamma)$ can be obtained by maximising $\ell_N(\bm\beta_0, \bm\beta_1, \bm\Sigma, \bm\gamma|\bm y , \bm\lambda)$ in \eqref{loglikN with AR(1)} with respect to $(\bm\beta_0, \bm\beta_1, \bm\Sigma, \bm\gamma)$.
This gives us a close form solution as follows:
\begin{align}
    \begin{pmatrix}\hat{\bm\beta_0'} \\ \hat{\bm\beta_1'} \\ \hat{\bm\gamma'}\end{pmatrix}
        =&
        \begin{pmatrix} S_{1/\lambda} & S'_{\bm x/\lambda} & n \\
                        S_{\bm x/\lambda} & S_{\bm x\bm x'/\lambda} & S_{\bm x} \\
                        n & S'_{\bm x} & S_\lambda\end{pmatrix}^{-1}
        \begin{pmatrix} S'_{\bm y/\lambda} \\ S_{\bm x\bm y'/\lambda} \\ S'_{\bm y} \end{pmatrix} \label{CM-step AR1 Beta} \\
    \hat{\bm\Sigma} \label{CM-step AR1 Sigma}
        =& \ \frac{1}{n} \sum_{i=1}^n \frac{1}{\lambda_i} (\bm y_i - \bm\beta_0 - \bm\beta_1 \bm y_{i-1})
                (\bm y_i - \bm\beta_0 - \bm\beta_1 \bm y_{i-1})'
            - \frac{1}{n} \bm\gamma\bm\gamma' S_\lambda,
\end{align}
where (\ref{CM-step AR1 Beta}) as a generalisation of (\ref{CM-step mu}) and (\ref{CM-step gamma}) can be further generalised to include higher order AR terms in the mean function. Then the complete data sufficient statistics are given by \eqref{suffstats} with the addition of
\begin{align} \label{suffstats AR(1)}
    S_{\bm x} = \sum_{i=1}^n \bm y_{i-1}, \
    S_{\bm x/\lambda} = \sum_{i=1}^n \frac{1}{\lambda_i} \bm y_{i-1}, \
    S_{\bm x \bm x'/\lambda} = \sum_{i=1}^n \frac{1}{\lambda_i} \bm y_{i-1} \bm y_{i-1}', \
    S_{\bm x \bm y'/\lambda} = \sum_{i=1}^n \frac{1}{\lambda_i} \bm y_{i-1} \bm y_{i}'.
\end{align} \par \vspace{3mm}

\noindent {\bf CM-step for $\nu$:} \\
This step is the same as in Section \ref{Subsection: Basic algorithm}. \vspace{3mm}

In summary, the layout of the MCECM and ECME algorithms are similar to those for the MSVG model in Section \ref{Subsection: Basic algorithm} except that we update $\bm\beta_0$ and $\bm\beta_1$ instead of $\bm\mu$ and adjust the E-step accordingly.

\subsection{Adjustment to ECM algorithm}    \label{Subsection: adjustment to ECM algorithm}

\subsubsection{Adjustment for small $\nu$}  \label{Subsubsection: adjustment for small v}
Numerical problems may occur when dealing with $\nu\leq\tfrac{d}{2}$ since the pdf $f_{VG}(\bm y_i)$ in \eqref{VGpdf} at $\bm\mu$ is unbounded for such $\nu$.
To handle such case, we can show that
as $\bm y_i \rightarrow \bm\mu$,
\begin{eqnarray*}
    \mathbb{E}(\lambda_i | \bm y_i, \bm \theta)
        & \sim &
        \begin{cases}
            \frac{2\nu-d}
                    {2\nu+\bm\gamma'\bm\Sigma^{-1}\bm\gamma}
                & \text{ if } \nu>\tfrac{d}{2}, \\
            -\frac{1}{(2\nu+\bm\gamma'\bm\Sigma^{-1}\bm\gamma)  \log \delta_i}
                & \text{ if } \nu=\tfrac{d}{2}, \\
            \frac{\Gamma(\nu-\tfrac{d}{2}+1)}{\Gamma(\tfrac{d}{2}-\nu)}
                    2^{2\nu - d + 1}
                    \left(2\nu+\bm\gamma'\bm\Sigma^{-1}\bm\gamma \right)^{\frac{d}{2}-\nu-1} \delta_i^{d-2\nu}
                & \text{ if } \nu<\tfrac{d}{2}, \\
        \end{cases} \\ \\ \\
    \mathbb{E}\left(\frac{1}{\lambda_i} \Bigg| \bm y_i, \bm \theta \right)
        & \sim &
        \begin{cases}
            \frac{2\nu+\bm\gamma'\bm\Sigma^{-1}\bm\gamma}{2\nu-d-2}
                & \text{ if } \nu>\tfrac{d}{2}+1, \\
            -\left(2\nu+\bm\gamma'\bm\Sigma^{-1}\bm\gamma\right) \log\delta_i
                & \text{ if } \nu=\tfrac{d}{2}+1, \\
            \frac{\Gamma(1-\nu+\tfrac{d}{2})}{\Gamma(\nu-\tfrac{d}{2})}
                    2^{1-2\nu+d} \left(2\nu+\bm\gamma'\bm\Sigma^{-1}\bm\gamma\right)^{\nu-\frac{d}{2}} \delta_i^{2\nu-d-2}
                & \text{ if } \nu\in(\tfrac{d}{2},\tfrac{d}{2}+1), \\
            -\frac{1}{\delta_i^2 \log\delta_i }
                & \text{ if } \nu=\tfrac{d}{2}, \\
            \frac{d-2\nu}{\delta_i^2 }
                & \text{ if } \nu<\tfrac{d}{2}, \\
        \end{cases} \\ \\ \\
    \mathbb{E}(\log\lambda_i | \bm y_i, \bm \theta)
        & \sim &
        \begin{cases}
            \psi(\nu-\frac{d}{2}) - \log\left(\frac{2\nu+\bm\gamma'\bm\Sigma^{-1}\bm\gamma}{2}\right)
                & \text{ if } \nu>\tfrac{d}{2}, \\
            \log\delta_i
                & \text{ if } \nu=\tfrac{d}{2}, \\
            2\log\delta_i
                & \text{ if } \nu<\tfrac{d}{2}.
        \end{cases}
\end{eqnarray*}

So for $\nu\leq\tfrac{d}{2}+1$ and $\nu\leq\tfrac{d}{2}$,
$\mathbb{E}(\frac{1}{\lambda_i} | \delta_i,\bm\gamma,\nu)$ and
$\mathbb{E}(\log\lambda_i | \delta_i, \bm\gamma,\nu)$ are unbounded for data points at the mean $\bm \mu$ respectively. As these expected values are numerically unstable to calculate, we propose to bound the pdf around $\bm\mu$ by a bound such that if
\begin{align}   \label{Delta region}
    \delta_i \sqrt{2\nu + \bm\gamma'\bm\Sigma^{-1}\bm\gamma} < \Delta ,
\end{align}
where $\Delta$ is a small fixed constant, then
we compute the expected values
    $\mathbb{E}(\frac{1}{\lambda_i} | \delta_i^*,\bm\gamma,\nu)$ and
    $\mathbb{E}(\log\lambda_i | \delta_i^*, \bm\gamma,\nu)$ in \eqref{E-step lambda2} and \eqref{E-step lambda3} with
    $\delta_i^* = \Delta/\sqrt{2\nu + \bm\gamma'\bm\Sigma^{-1}\bm\gamma}$ replacing $\delta_i$.
The region in \eqref{Delta region} will be called the delta region.
We will analyse the optimal choices of $\Delta$ in a simulation study in the next section.

Another problem may occur when estimating $\bm\Sigma$.
Suppose that $\bm y_i \rightarrow \bm\mu^{(t)}$ for some $i$, then $\widehat{1/\lambda_i}^{(t+1/2)} \rightarrow \infty$.
Now consider the CM-step for $(\bm\mu,\bm\Sigma,\bm\gamma)$, $\bm\mu^{(t+1)}$ needs to be calculated before calculating $\bm\Sigma^{(t+1)}$.
For the $i\textsuperscript{th}$ term of the first summation for calculating $\bm\Sigma^{(t+1)}$ in \eqref{CM-step Sigma}, we get that
\begin{align*}
    \widehat{1/\lambda_i}^{(t+1/2)} (\bm y_i - \bm\mu^{(t+1)})(\bm y_i - \bm\mu^{(t+1)})' \rightarrow \infty ,
\end{align*}
which leads to $\bm\Sigma^{(t+1)}$ diverging to infinity since $(\bm y_i - \bm\mu^{(t+1)})$ does not converge to zero.
To solve this problem, we insert an extra E-step to update $\widehat{1/\lambda_i}^{(t+3/4)}$ and $\widehat{\lambda}_i^{(t+3/4)}$ after updating $(\bm\mu^{(t+1)},\bm\gamma^{(t+1)})$. This adjustment makes the $i\textsuperscript{th}$ term of the first summation in $\bm\Sigma^{(t+1)}$ to be
\begin{align*}
    \widehat{1/\lambda_i}^{(t+3/4)} (\bm y_i - \hat{\bm\mu}^{(t+1)})(\bm y_i - \hat{\bm\mu}^{(t+1)})' \rightarrow \bm C
\end{align*}
for some finite constant matrix $\bm C$ resulting in a more numerically stable estimate for $\bm\Sigma^{(t+1)}$.
Thus adding an additional E-step to update $\widehat{1/\lambda_j}^{(t+3/4)}$ after updating $\bm\mu^{(t+1)}$ and $\bm\gamma^{(t+1)}$ before updating $\bm\Sigma^{(t+1)}$ improves the numerical stability and convergence of the algorithm.
\par \vspace{3mm}

\subsubsection{Hybrid ECM algorithm}        \label{Subsubsection: Hybrid ECM algorithm}

\citet{Meng1993}, \citet{Liu1994} together showed that the MCECM and ECME algorithm always increase the log-likelihood monotonically.
However, problems might occur when dealing with multimodel log-likelihood such as the MSVG log-likelihood when $\nu\leq\frac{d}{2}$.

In the MCECM algorithm, $\nu$ is estimated by numerically maximising the log-likelihood of the conditional gamma mixing density using NR algorithm with derivatives \eqref{NR1} and \eqref{NR2}.
The ECME algorithm which numerically maximises the actual log-likelihood of the MSVG distribution with pdf \eqref{VGpdf}, is computationally less efficient than the MCECM algorithm despite less iterations are required for convergence.
However, the ECME algorithm is more stable than the MCECM algorithm since the CM-step for $\nu$ maximises the actual log-likelihood, hence avoiding the missing data $\widehat{\log \lambda}_i$ in CM-step 2.

In light of this, we propose an algorithm which combines the efficiency of the MCECM algorithm while maintaining the performance of the ECME algorithm.
We call this the {\it hybrid ECM} (HECM) algorithm.
Essentially it starts with the MCECM algorithm and repeats until either the relative increment of log-likelihood is smaller than $10^{-8}$
which stops the algorithm, reverts back to the previous estimates and performs the ECME algorithm.

For the univariate skew VG model, the performance of the HECM algorithm can be improved by replacing the estimate for $\hat\mu$ in \eqref{CM-step mu} with the estimate for $\mu$ such that it maximizes the actual log-likelihood $\ell_{VG}(\mu|\bm y, \sigma^2, \gamma, \nu)$.

The two adjustments using density bound and extra E-step as well as the hybrid procedures balance computational efficiency, and accuracy of the ML estimation.

\subsection{Observed Information Matrix}

Letting $\bm Y_c=(\bm y, \bm \lambda)$ be the complete data,
\cite{Louis1982} expressed the observed information matrix $\bm I_o(\bm\theta)$ in terms of the conditional expectation of the derivatives of complete data log-likelihood $\ell'(\bm \theta| \bm Y_c)$ and $\ell''(\bm \theta| \bm Y_c)$
with respect to conditional distribution $\bm\lambda|\bm y$ which depends on $\bm\theta$:
\begin{equation} \label{EMse}
    \bm I_o(\bm\theta)
        = -\mathbb E_{\bm\theta} [\ell''(\bm \theta| \bm Y_c)]
            - \mathbb E_{\bm\theta} [\ell'(\bm \theta| \bm Y_c) \ell'^T(\bm \theta| \bm Y_c)]
            + \mathbb E_{\bm\theta} [\ell'(\bm \theta| \bm Y_c)]
                \mathbb E_{\bm\theta} [\ell'(\bm \theta| \bm Y_c)]^T.
\end{equation}
In the context of MSVG model, $\ell(\bm \theta| \bm Y_c)$ is given by \eqref{decomposed loglik}.
The first derivatives of complete data log-likelihood can be expressed as
\begin{align*}
    &\frac{\partial\ell_N}{\partial\bm\beta_0}
        = \bm\Sigma^{-1} \sum_{i=1}^n \frac{1}{\lambda_i} (\bm y_i - \bm\beta_0 - \bm\beta_1 \bm y_{i-1} - \lambda_i\bm\gamma) \\
    &\frac{\partial\ell_N}{\partial\text{vec}\bm\beta_1}
        = \text{vec}\left[\bm\Sigma^{-1} \sum_{i=1}^n \frac{1}{\lambda_i} (\bm y_i - \bm\beta_0 - \bm\beta_1 \bm y_{i-1} - \lambda_i\bm\gamma) \bm y_{i-1}'\right] \\
    &\frac{\partial\ell_N}{\partial\text{vech}\bm\Sigma}
        = \text{vech}\left(2(\bm1_{p\times p}) - \bm I\right)\circ
            \text{vech}\left(-\frac{n}{2}\bm\Sigma^{-1}
                + \frac{1}{2}\bm\Sigma^{-1} S_{\tilde{\bm y} \tilde{\bm y}'/\lambda} \bm\Sigma^{-1}\right) \\
    &\frac{\partial\ell_N}{\partial\bm\gamma}
        = \bm\Sigma^{-1} \sum_{i=1}^n (\bm y_i - \bm\beta_0 - \bm\beta_1\bm y_{i-1} - \lambda_i\bm\gamma) \\
    &\frac{\partial\ell_G}{\partial\nu} = n + n\log\nu -n\psi(\nu) + \sum_{i=1}^n \log\lambda_i - \sum_{i=1}^n \lambda_i
\end{align*}
where $\text{vec}(\cdot)$ and $\text{vech}(\cdot)$ are the vectorisation and half-vectorisation operators respectively,
$\bm1_{p\times p}$ is the $p \times p$ matrix of 1's,
$\bm I$ be a conformable identity matrix, and
\begin{align*}
    S_{\tilde{\bm y} \tilde{\bm y}'/\lambda}
        =& \sum_{i=1}^n \frac{1}{\lambda_i} (\bm y_i - \bm\beta_0 - \bm\beta_1 \bm y_{i-1} - \lambda_i \bm\gamma)(\bm y_i - \bm\beta_0 - \bm\beta_1 \bm y_{i-1} - \lambda_i \bm\gamma)'.
\end{align*}

Calculating the conditional expectations
in equation (\ref{EMse})
requires the second order derivatives which can be obtained using vector differentiation, and
the following conditional expectations
\begin{align*}
    \mathbb E_{\bm\theta}\left(\lambda_i^k\right)
        =& \left(\frac{\delta_i}{\psi}\right)^k
            \frac{K_{\nu-\frac{d}{2}+k}\left(\delta_i\psi\right)}
                {K_{\nu-\frac{d}{2}}\left(\delta_i\psi\right)} \\
    \mathbb E_{\bm\theta}\left(\lambda_i^k\log\lambda_i\right)
        =& \left(\frac{\delta_i}{\psi }\right)^k
            \frac{1}{K_{\nu-\frac{d}{2} }\left(\delta_i\psi\right)}
                \left[K^{(1,0)}_{\nu-\frac{d}{2}+k}\left(\delta_i\psi\right)+\log \left(\frac{\delta_i}{\psi}\right) K_{\nu-\frac{d}{2}+k}\left(\delta_i\psi\right)\right] \\
    \mathbb E_{\bm\theta}\left((\log\lambda_i)^2\right)
        =& \left[\log \left(\frac{\delta_i}{\psi }\right)\right]^2
            +\frac{K_{\nu-\frac{d}{2} }^{(2,0)}\left(\delta_i\psi\right)
                    +2\log \left(\frac{\delta_i}{\psi }\right) K_{\nu-\frac{d}{2} }^{(1,0)}\left(\delta_i\psi\right)}
                {K_{\nu-\frac{d}{2} }\left(\delta_i\psi\right)}
\end{align*}
where we let $\psi=\sqrt{\bm\gamma'\bm\Sigma^{-1}\bm\gamma+2\nu}$ and
$K_{\nu}^{(2,0)}(z) = \tfrac{\partial^2}{\partial\alpha^2} K_\alpha (z)\big|_{\alpha=\nu}$ which is approximated using central difference formula.

The asymptotic covariance matrix of $\hat{\bm\theta}$ can be approximated by the inverse of the observed information matrix $\bm I_o(\hat{\bm\theta})$.
This gives us a way to approximate the standard error of $\hat\theta_i=(\hat{\bm\theta})_i$
\begin{equation}
    SE\left(\hat{\theta_i}\right)
        \approx \sqrt{\left[\bm I_o(\hat{\bm\theta})^{-1}\right]_{ii}} \ .
\end{equation}

\section{Simulation Study}      \label{Section: Simulation study}

We perform three simulation studies:
    firstly, to compare the efficiency of the MCECM, ECME and HECM algorithms with the extra E-step mentioned in Section \ref{Subsubsection: adjustment for small v};
    secondly, to analyse the optimal choices of $\Delta$ on increasing the dimension of MSVG distribution; and
    lastly, to assess the performance of HECM algorithm for different levels of shape and skewness parameters.
We use the statistical program called \texttt R to generate $r$ random samples from the MSVG distribution each of sample size $n=1000$
and estimate the parameters for each sample.
We set the true parameters to be
\begin{equation*}
    \bm\mu=
        \begin{pmatrix}
            0 \\
            0
        \end{pmatrix} , \quad
    \bm\Sigma=
        \begin{pmatrix}
            1 & 0.4 \\
            0.4 & 1
        \end{pmatrix} , \quad
    \bm\gamma=
        \begin{pmatrix}
            0.2 \\
            0.3
        \end{pmatrix} , \quad
    \nu=3
\end{equation*}
for the bivariate base model and use initial values recommended in Section \ref{Subsection: Basic algorithm},
that is $(\bm\mu_0, \bm\Sigma_0, \bm\gamma_0, \nu_0) = (\bar {\bm y}, \bm Q_{y},\bm 0,2)$.
To improve convergence, we multiply the data by $C=100$.
Thus we have the scaled parameters as
    $C \bm \mu$,
    $C^2 \bm \Sigma$ and
    $C \bm \gamma$ for $\bm \mu$, $\bm \Sigma$ and $\bm \gamma$ respectively.
The results are then rescaled back and
for each random sample, we calculate the log-likelihood to assess the model fit.

\subsection{Efficiency across algorithms}
To illustrate that the three algorithms, MCECM, ECME, and HECM provide good estimates, we perform the simulation above with $r=1000$ replications using the bivariate base model and compare their relative performance.
In Table \ref{analysing efficiency of algorithms} below, the average of parameter estimates, log-likelihoods, number of iterations for convergence over replications, and the total computational times are presented.
For each algorithm, we generate the same set of simulated data using the same seed to eliminate the effect of sampling errors in the comparison.

\begin{table}[H]
\caption{Parameter estimates and model efficiency measures across algorithms}
    \begin{center}
    \begin{tabular}{|l|r|r|r|r|}
        \hline
         & \multicolumn{1}{l|}{true} & \multicolumn{1}{c|}{MCECM} & \multicolumn{1}{c|}{ECME} & \multicolumn{1}{c|}{HECM} \\
            \hline\hline
        $\mu_1$ & 0 & $-$0.0072 & $-$0.0072 & $-$0.0072 \\
        $\mu_2$ & 0 & $-$0.0069 & $-$0.0069 & $-$0.0069 \\
            \hline
        $\Sigma_{1,1}$ & 1 & 0.9959 & 0.9959 & 0.9959 \\
        $\Sigma_{1,2}$ & 0.4 & 0.3973 & 0.3973 & 0.3973 \\
        $\Sigma_{2,2}$ & 1 & 0.9914 & 0.9914 & 0.9914 \\
            \hline
        $\gamma_1$ & 0.2 & 0.2067 & 0.2067 & 0.2068 \\
        $\gamma_2$ & 0.3 & 0.3062 & 0.3062 & 0.3062 \\
            \hline
        $\nu$ & 2.5 & 2.5699 & 2.5709 & 2.5710 \\
            \hline \hline
        loglik & \multicolumn{1}{l|}{} & $-$2713.41 & $-$2713.41 & $-$2713.41 \\
        conv.iter & \multicolumn{1}{l|}{} & 75.6 & 31.8 & 76.6 \\
        time (hr) & \multicolumn{1}{l|}{} & 8.1 & 16.4 & 9.1 \\ \hline
    \end{tabular}
    \end{center}
\label{analysing efficiency of algorithms}
\end{table}

All estimates obtained are reasonably close to the true value which supports the validity of these algorithms.
As expected, MCECM algorithm is the most efficient in terms of computation time whereas ECME algorithm converges the fastest in terms of the number of iterations.
However the ECME and HECM algorithms provide slightly better model fits when comparing more digits of the averaged log-likelihood.
Overall HECM algorithm is preferred as it provides better model fit than the MCECM algorithm and runs faster than the ECME algorithm. Hence HECM algorithm will be adopted in all subsequent analyses.

\subsection{Efficiency across $\Delta$ for small shape parameter} \label{Section: analysing delta}
In this simulation study, we analyse the behavior of estimates for different levels of $\Delta$ as defined in Section \ref{Subsubsection: adjustment for small v}.
We begin the simulation study with bivariate skewed VG distribution with parameters
as in Table \ref{analysing efficiency of algorithms} but with $\nu=0.6$ as the true shape parameter.
The shape parameter is chosen to satisfy the condition $\nu\leq\tfrac{d}{2}$ for unbounded density. We repeat the experiment for different levels of $\Delta$ but we do not report average log-likelihood because it can be estimated to be as large as possible by reducing the value of $\Delta$.
For each $\Delta$ level, we repeat the estimation procedure $r=1000$ times.
Generally from the results of the experiment as shown in Table \ref{analysing delta 2dim},
the further away the $\Delta$ is from the range (1e-5,1e-3), the further away are the estimates from their true values.
Hence, an optimal range of $\Delta$ is from 1e-5 to 1e-3.

This experiment is repeated for dimension $d=3$ with $\nu=1$.
The results for these experiments are given in Tables \ref{analysing delta 3dim}.

\begin{table}[H]
\caption{Parameter estimates across levels of $\Delta$ for skewed bivariate VG model.}
\begin{center}
\begin{tabular}{|l|r|r|r|r|r|r|r|r|r|}
    \hline
 & & \multicolumn{8}{|c|}{$\Delta$ levels}  \\ \cline{3-10}
 & true & \multicolumn{1}{c|}{1e-300} & \multicolumn{1}{c|}{1e-100} & \multicolumn{1}{c|}{1e-50} & \multicolumn{1}{c|}{1e-10} & \multicolumn{1}{c|}{1e-7} & \multicolumn{1}{c|}{1e-5} & \multicolumn{1}{c|}{1e-4} & \multicolumn{1}{c|}{1e-3} \\
    \hline \hline
$\mu_1$ & 0 & 0.0040 & 0.0040 & 0.0040 & 0.0040 & 0.0040 & 0.0040 & 0.0040 & \blue{\textbf{0.0039}} \\
$\mu_2$ & 0 & 0.0058 & 0.0058 & 0.0058 & 0.0058 & 0.0058 & 0.0058 & 0.0058 & \blue{\textbf{0.0056}} \\
    \hline
$\Sigma_{1,1}$ & 1 & 1.2118 & 1.0600 & 1.0434 & 1.0111 & 1.0043 & \blue{\textbf{0.9997}} & 0.9961 & 0.9916 \\
$\Sigma_{1,2}$ & 0.4 & 0.4880 & 0.4251 & 0.4184 & 0.4054 & 0.4027 & 0.4009 & \blue{\textbf{0.3994}} & 0.3976 \\
$\Sigma_{2,2}$ & 1 & 1.2149 & 1.0607 & 1.0438 & 1.0116 & 1.0049 & \blue{\textbf{1.0003}} & 0.9967 & 0.9921 \\
    \hline
$\gamma_1$ & 0.2 & 0.1939 & 0.1910 & 0.1932 & 0.1952 & 0.1952 & 0.1953 & 0.1951 & \blue{\textbf{0.1950}} \\
$\gamma_2$ & 0.3 & 0.2904 & 0.2861 & 0.2894 & 0.2925 & 0.2925 & \blue{\textbf{0.2925}} & 0.2923 & 0.2921 \\
    \hline
$\nu$ & 0.6 & 0.3300 & 0.4421 & 0.4964 & 0.5771 & 0.5885 & 0.5965 & \blue{\textbf{0.6010}} & 0.6053 \\
    \hline
    \multicolumn{6} {l} {\footnotesize Estimate in \textbf{bold} is closest to the true value.}
\end{tabular}
\end{center}
\label{analysing delta 2dim}
\end{table}

\begin{table}[H]
\caption{Parameter estimates across levels of $\Delta$ for skewed trivariate VG model.}
\begin{center}
\begin{tabular}{|l|r|r|r|r|r|r|r|r|}
    \hline
 & & \multicolumn{7}{|c|}{$\Delta$ levels}  \\ \cline{3-9}
 & true & \multicolumn{1}{c|}{1e-7} & \multicolumn{1}{c|}{1e-6} & \multicolumn{1}{c|}{1e-5} & \multicolumn{1}{c|}{1e-4} & \multicolumn{1}{c|}{1e-3} & \multicolumn{1}{c|}{1e-2} & \multicolumn{1}{c|}{1e-1} \\
    \hline \hline
$\mu_1$ & 0 & -0.0019 & -0.0019 & -0.0019 & -0.0019 & -0.0019 & -0.0018 & \blue{\textbf{0.0013}} \\
$\mu_2$ & 0 & -0.0024 & -0.0024 & -0.0024 & -0.0024 & -0.0024 & -0.0022 & \blue{\textbf{-0.0020}} \\
$\mu_3$ & 0 & -0.0012 & -0.0012 & -0.0012 & -0.0012 & -0.0012 & \blue{\textbf{-0.0011}} & -0.0015 \\
    \hline
$\Sigma_{1,1}$ & 1 & 1.0092 & 1.0070 & 1.0049 & 1.0026 & \blue{\textbf{0.9989}} & 0.9952 & 0.9889 \\
$\Sigma_{1,2}$ & 0.4 & 0.4020 & 0.4012 & \blue{\textbf{0.4003}} & 0.3994 & 0.3980 & 0.3966 & 0.3942 \\
$\Sigma_{1,3}$ & 0.3 & 0.3019 & 0.3013 & 0.3007 & \blue{\textbf{0.3000}} & 0.2989 & 0.2978 & 0.2960 \\
$\Sigma_{2,2}$ & 1 & 1.0094 & 1.0073 & 1.0051 & 1.0029 & \blue{\textbf{0.9992}} & 0.9956 & 0.9893 \\
$\Sigma_{2,3}$ & 0.2 & 0.2007 & 0.2003 & \blue{\textbf{0.1999}} & 0.1995 & 0.1988 & 0.1981 & 0.1969 \\
$\Sigma_{3,3}$ & 1 & 1.0128 & 1.0107 & 1.0086 & 1.0064 & 1.0027 & \blue{\textbf{0.9991}} & 0.9926 \\
    \hline
$\gamma_1$ & 0.2 & 0.2012 & 0.2012 & 0.2012 & 0.2012 & 0.2010 & 0.2007 & \blue{\textbf{0.1999}} \\
$\gamma_2$ & 0.3 & 0.3026 & 0.3026 & 0.3026 & 0.3026 & 0.3025 & 0.3019 & \blue{\textbf{0.3013}} \\
$\gamma_3$ & 0.4 & 0.4018 & 0.4018 & 0.4018 & 0.4018 & 0.4016 & 0.4010 & \blue{\textbf{0.4009}} \\
    \hline
$\nu$ & 1 & 0.9512 & 0.9602 & 0.9696 & 0.9795 & 0.9912 & \blue{\textbf{1.0013}} & 1.0225 \\
    \hline
    \multicolumn{6} {l} {\footnotesize Estimate in \textbf{bold} is closest to the true value.}
\end{tabular}
\end{center}
\label{analysing delta 3dim}
\end{table}

On increasing the dimension, the optimal ranges for $\Delta$ is (1e-5,1e-1).
As for the subsequent simulation experiment, we implement $\Delta$ within the optimal ranges for dimension 2 to 3. As results show that the optimal $\Delta$ range only increases slowly with $d$, we apply the optimum $\Delta$ range for $d=3$ to the five dimensional MSVG model in the real data analysis.

For the univariate case, we use the modified HECM algorithm as mentioned in Section \ref{Subsubsection: Hybrid ECM algorithm}. The results are presented below in Table \ref{analysing delta 1dim}.
Due to the modification, it is not possible to directly compare the algorithm for high dimensional cases.
As the $\Delta$ level gets bigger, $\hat\sigma^2$ and $\hat\gamma$ increases towards the true value. However, $\hat\nu$ increases away from the true value.
Since the optimal $\Delta$ range is not clear for the univariate case, we suggest any $\Delta$ in the range (1e-10,1e-3).

\begin{table}[H]
    \caption{Parameter estimates across levels of $\Delta$ for symmetric univariate VG model}
    \begin{center}
    \begin{tabular}{|l|r|r|r|r|r|r|r|r|r|}
        \hline
     & & \multicolumn{7}{|c|}{$\Delta$ levels}  \\ \cline{3-9}
     & true & \multicolumn{1}{c|}{1e-300} & \multicolumn{1}{c|}{1e-100} & \multicolumn{1}{c|}{1e-50} & \multicolumn{1}{c|}{1e-20} & \multicolumn{1}{c|}{1e-10} & \multicolumn{1}{c|}{1e-5} & \multicolumn{1}{c|}{1e-3} \\
        \hline \hline
    $\mu$ & 0 & 0.0000 & 0.0000 & 0.0000 & 0.0000 & 0.0000 & 0.0000 & 0.0000 \\
    $\sigma^2$ & 1 & 0.8564 & 0.8564 & 0.8564 & 0.8564 & 0.8663 & 0.8862 & 0.9148 \\
    $\gamma$ & 0.2 & 0.1878 & 0.1878 & 0.1878 & 0.1878 & 0.1885 & 0.1907 & 0.1952 \\
    $\nu$ & 0.3 & 0.3102 & 0.3102 & 0.3102 & 0.3102 & 0.3111 & 0.3137 & 0.3267 \\
        \hline
    \end{tabular}
    \end{center}
    \label{analysing delta 1dim}
\end{table}

\subsection{Effects of shape and skewness parameters}
In this simulation study, we study the performance of HECM algorithm
for various levels of skewness parameter $\bm\gamma$ when the shape parameter is $\nu=3$ and $\nu=0.6$ which indicates large and small value respectively according to the condition $\nu\leq\tfrac{d}{2}$.
We consider the same bivariate base model as described at the start of Section \ref{Section: Simulation study} except that
the true skewness parameters are
$\bm\gamma$ = (0.2,0.2), (0.5,0.5), (0.7,0.7), (1,1), (2,2) and (0.5,2).
We report the averaged iteration in which the algorithm switches from MCECM to ECME algorithm as denoted by \textit{switch iter}, the averaged iteration when the algorithm converges by \textit{conv iter}, and the total time it takes to run the algorithm.

\begin{table}[H]
\caption{Parameter estimates and model efficiency measures for $\nu=3$}
\begin{center}
\begin{tabular}{|l|l|r|r|r|r|r|r|}
    \hline
 & & \multicolumn{6}{|c|}{True $(\gamma_1,\gamma_2)$}  \\ \cline{3-8}
 & true & \multicolumn{1}{c|}{$(0.2,0.2)$} & \multicolumn{1}{c|}{$(0.5,0.5)$} & \multicolumn{1}{c|}{$(0.7,0.7)$} & \multicolumn{1}{c|}{$(1,1)$} & \multicolumn{1}{c|}{$(2,2)$} & \multicolumn{1}{c|}{$(0.5,2)$} \\
    \hline \hline
$\mu_1$ & 0 & -0.0081 & -0.0129 & -0.0149 & -0.0176 & -0.0246 & -0.0071 \\
$\mu_2$ & 0 & -0.0095 & -0.0143 & -0.0162 & -0.0186 & -0.0244 & -0.0292 \\
    \hline
$\Sigma_{1,1}$ & 1 & 0.9960 & 0.9942 & 0.9927 & 0.9903 & 0.9779 & 0.9951 \\
$\Sigma_{1,2}$ & 0.4 & 0.3985 & 0.3963 & 0.3947 & 0.3921 & 0.3790 & 0.3932 \\
$\Sigma_{2,2}$ & 1 & 0.9952 & 0.9931 & 0.9915 & 0.9888 & 0.9759 & 0.9727 \\
    \hline
$\gamma_1$ & & 0.2061 & 0.5108 & 0.7128 & 1.0154 & 2.0221 & 0.5050 \\
$\gamma_2$ & & 0.2084 & 0.5131 & 0.7150 & 1.0173 & 2.0229 & 2.0277 \\
    \hline
$\nu$ & 3 & 3.1006 & 3.0856 & 3.0785 & 3.0699 & 3.0477 & 3.0535 \\
    \hline \hline
switch.iter & & 99.31 & 105.34 & 111.89 & 125.73 & 241.16 & 192.61 \\
conv.iter & & 99.31 & 105.34 & 111.89 & 125.73 & 241.16 & 192.61 \\
time (hr) & & 11.5 & 12.1 & 12.3 & 22.8 & 41.2 & 34.2 \\
    \hline
\end{tabular}
\end{center}
\label{analysing skewness df3}
\end{table}

Table \ref{analysing skewness df3} shows that all estimates are very close to their true values. On increasing the level of skewness, the biases of parameters $\bm\mu$, $\bm\gamma$ and $\bm\Sigma$ tend to increase gradually while the bias for $\nu$ decreases gradually.
The number of iterations to switch and converge and hence the computation time also increase accordingly.
This also applies for the unequal skewness case.
Moreover, there is no reliance of ECME algorithm as the parameter $\nu$ is chosen so that $\nu > d/2$.
Indeed the simulation with larger skewness needs more iterations for the algorithm to converge.
The results when $\nu=0.6$ is presented in Table \ref{analysing skewness df0.6}.

\begin{table}[H]
\caption{Parameter estimates and model efficiency measures for $\nu=0.6$}
\begin{center}
\begin{tabular}{|l|l|r|r|r|r|r|r|}
    \hline
 & & \multicolumn{6}{|c|}{True $(\gamma_1,\gamma_2)$}  \\ \cline{3-8}
 & true & \multicolumn{1}{l|}{$(0.2,0.2)$} & \multicolumn{1}{l|}{$(0.5,0.5)$} & \multicolumn{1}{l|}{$(0.7,0.7)$} & \multicolumn{1}{c|}{$(1,1)$} & \multicolumn{1}{c|}{$(2,2)$} & \multicolumn{1}{l|}{$(0.5,2)$} \\
    \hline \hline
$\mu_1$ & 0 & 0.0043 & 0.0083 & 0.0108 & 0.0136 & 0.0212 & 0.0052 \\
$\mu_2$ & 0 & 0.0042 & 0.0082 & 0.0110 & 0.0138 & 0.0213 & 0.0213 \\
    \hline
$\Sigma_{1,1}$ & 1 & 0.9995 & 1.0054 & 1.0114 & 1.0220 & 1.0573 & 1.0175 \\
$\Sigma_{1,2}$ & 0.4 & 0.4006 & 0.4061 & 0.4124 & 0.4226 & 0.4391 & 0.4167 \\
$\Sigma_{2,2}$ & 1 & 0.9990 & 1.0044 & 1.0111 & 1.0217 & 1.0540 & 1.0558 \\
    \hline
$\gamma_1$ & & 0.1950 & 0.4904 & 0.6876 & 0.9845 & 2.0337 & 0.5007 \\
$\gamma_2$ & & 0.1943 & 0.4897 & 0.6866 & 0.9835 & 2.0328 & 2.0028 \\
    \hline
$\nu$ & 0.6 & 0.5968 & 0.5961 & 0.5957 & 0.5945 & 0.5946 & 0.5940 \\
    \hline \hline
switch.iter &  & 26.8 & 27.6 & 28.1 & 23.7 & 23.2 & 26.5 \\
conv.iter &  & 28.4 & 28.0 & 28.5 & 24.5 & 23.2 & 26.5 \\
time (hr) & & 4.9 & 3.5 & 3.7 & 6.7 & 5.8 & 6.6 \\
    \hline
\end{tabular}
\end{center}
\label{analysing skewness df0.6}
\end{table}

The parameter estimates using HECM algorithm and optimal $\Delta$ to deal with unbounded density are reasonably accurate.
This justifies the choice of $\Delta$ for $d=2$.
Surprisingly the algorithm roughly needs around the same number of iterations for each skewness level.
In comparison with the results in Table \ref{analysing skewness df3} for larger $\nu$, the algorithm requires significantly less iterations.
Eventually, the HECM algorithm switch to ECME to obtain parameter estimates.
This justifies our proposed HECM algorithm instead of relying solely on MCECM. 
\par \vspace{3mm}
In summary, the proposed HECM algorithm gives good parameter estimates for the MSVG distribution even when its shape parameter is small and skewness is heavy.
\par \vspace{3mm}

\section{Real Data Analysis}    \label{Section: Real data analysis}
\begin{figure}[htbp]
  \begin{center}
    \subfigure{\label{}\includegraphics[width=0.32\linewidth]{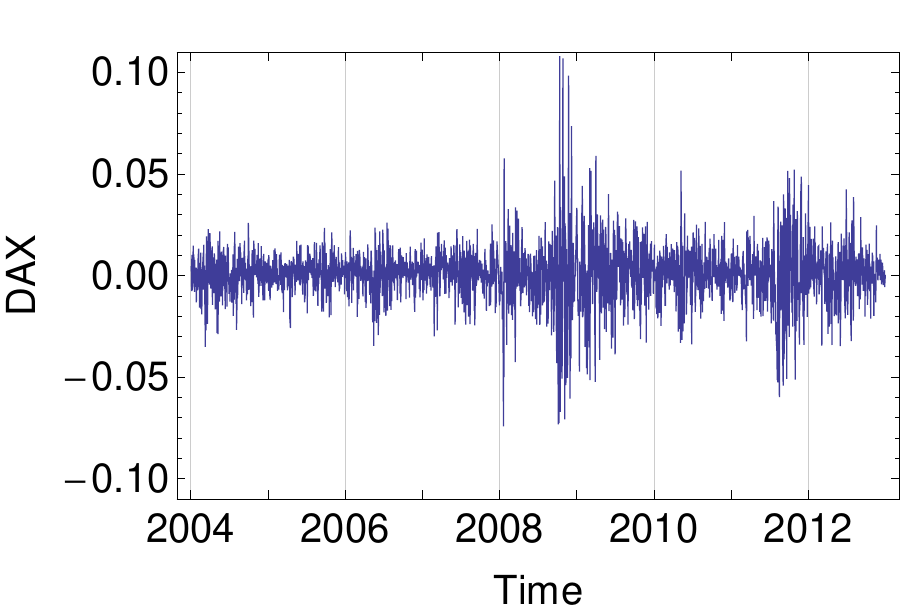}}
    \subfigure{\label{}\includegraphics[width=0.32\linewidth]{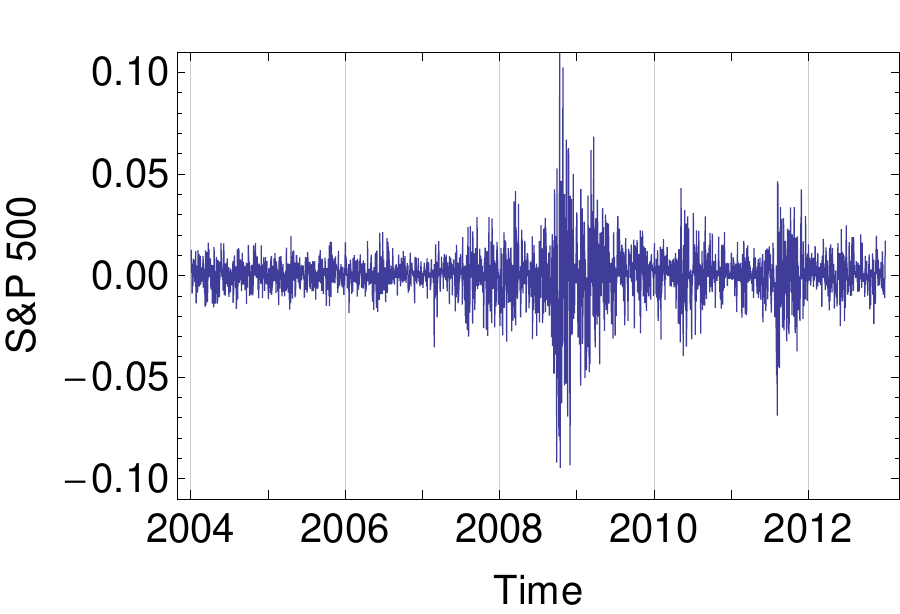}}
    \subfigure{\label{}\includegraphics[width=0.32\linewidth]{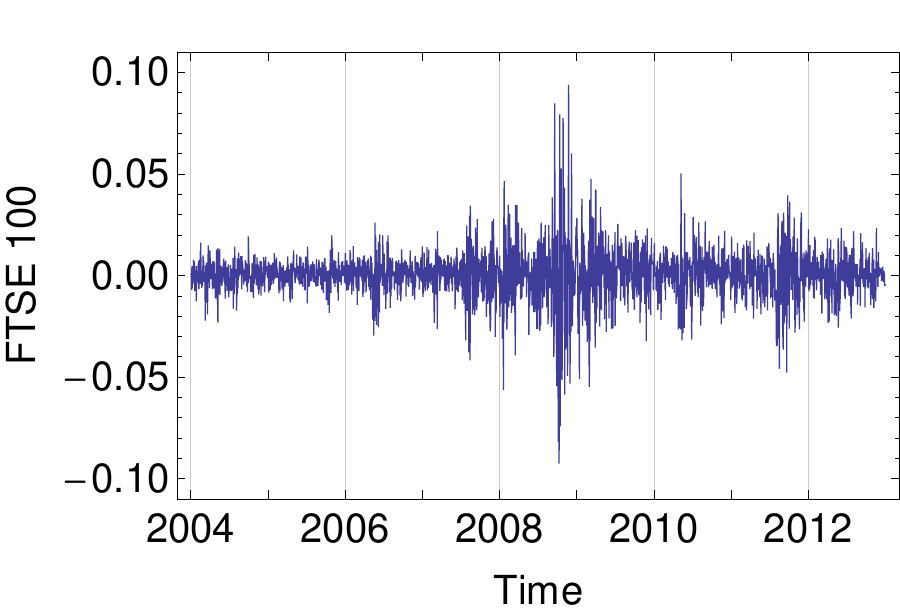}}
    \subfigure{\label{}\includegraphics[width=0.32\linewidth]{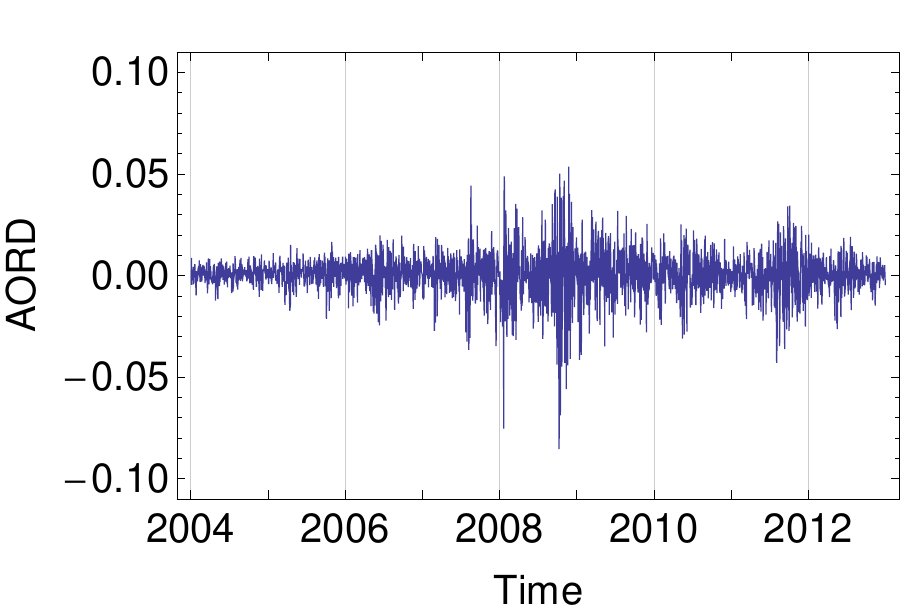}}
    \subfigure{\label{}\includegraphics[width=0.32\linewidth]{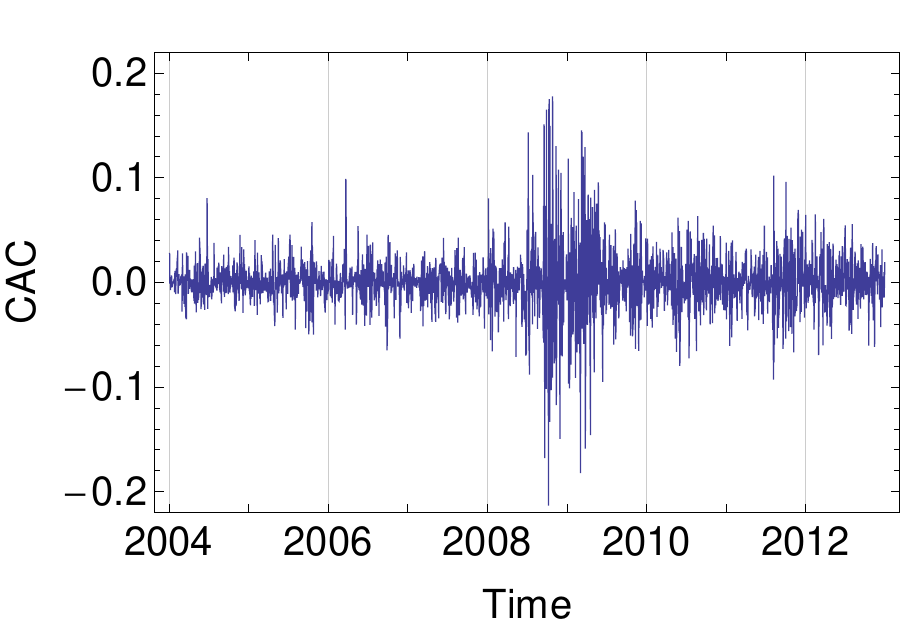}}
  \end{center}
  \caption{Time series plots for the five daily return series}
  \label{time series plots}
\end{figure}

To illustrate the applicability of HECM algorithm using the MSVG AR model,
we consider the returns of five daily closing price indices, namely, Deutscher Aktien (DAX), Standard \& Poors 500 (S\&P 500), Financial Times Stock Exchange 100 (FTSE 100), All Ordinaries (AORD) and Cotation Assiste en Continu 40 (CAC) from 1\textsuperscript{st} January 2004 to 31\textsuperscript{st} December 2012 with tolerance level of $10^{-10}$.
After filtering the data with missing closing prices, we obtain the data size of $n=2188$. Plots of the five time series are given in Figure \ref{time series plots}. As the summary statistics in Table \ref{summary statistics} show that the data exhibit considerable skewness and kurtosis, we begin our analyses with the MSVG model adopting a constant mean.
To assess the model fit, we use the Akaike information criterion with a correction for finite sample size (AICc) given by:
\begin{equation*}
    AICc = AIC + \frac{2k(k+1)}{n-k-1}
\end{equation*}
where $k$ denotes the number of model parameters. Model with the lowest AICc value is preferred as it demonstrates the best model fit after accounting for model complicity. \par \vspace{3mm}

\begin{table}[H]
\caption{Summary statistics for the five daily return series}
\begin{center}
\begin{tabular} {|l|r|r|r|r|r|} \hline
& \multicolumn{1}{c|}{DAX} & \multicolumn{1}{c|}{S\&P 500} & \multicolumn{1}{c|}{FTSE 100} & \multicolumn{1}{c|}{AORD} & \multicolumn{1}{c|}{CAC} \\ \hline \hline
mean$^{\dag}$  &  0.2901 & 0.1096 & 0.1223 & 0.1587 & 0.1755 \\ \hline
sd   &   0.0146  &  0.0136  &  0.0127  &  0.0114  &  0.0287 \\ \hline
max  &   0.1080  &  0.1096  &  0.0938  &  0.0540  &  0.1778 \\ \hline
min  &  $-$0.0774  & $-$0.0947  & $-$0.0926  & $-$0.1049  & $-$0.2133 \\ \hline
skewness  &  0.0208 &  $-$0.2939 &  $-$0.0946 &  $-$0.7291 &  0.1524 \\ \hline
kurtosis  &  9.4529 &  12.9395 &  11.0138 &  10.4846 &  11.7483 \\ \hline
\multicolumn{6} {l} {\footnotesize $\dag$ The reported means are multiplied by 1000.}
\end{tabular}
\end{center}
\label{summary statistics}
\end{table}

The results for the estimated parameters and their standard errors are given in Table \ref{msVG results} .

\begin{table}[H]
\caption{Results for the Multivariate Skewed VG model}
\begin{tabular}{|l|c|c|}
    \hline
 & estimate & standard error \\
    \hline \hline
$\bm\mu^T$ &
        $10^{-4}\begin{pmatrix}
            18 & 10 & 12 & 20 & -7
        \end{pmatrix}$ &
        $10^{-4}\begin{pmatrix}
            3 & 3 & 3 & 3 & 6
        \end{pmatrix}$ \\
    \hline
$\bm\Sigma$ &
        $10^{-5}\begin{pmatrix}
            19 & 10 & 14 & 5 & 12 \\
             & 14 & 8 & 2 & 17 \\
             &  & 13 & 4 & 9 \\
             &  &  & 12 & 0.4 \\
             &  &  &  & 66
        \end{pmatrix}$ &
        $10^{-6}\begin{pmatrix}
            7 & 5 & 6 & 4 & 9 \\
             & 5 & 4 & 3 & 9 \\
             &  & 5 & 3 & 7 \\
             &  &  & 5 & 7 \\
             &  &  &  & 26
        \end{pmatrix}$
         \\
    \hline
$\bm\gamma^T$ &
        $10^{-4}\begin{pmatrix}
           -15 & -9 & -11 & -19 & 9
        \end{pmatrix}$ &
        $10^{-4}\begin{pmatrix}
            5 & 4 & 4 & 4 & 8
        \end{pmatrix}$
         \\
    \hline
$\nu$ & 1.40 & 0.054 \\
    \hline \hline
Corr &
        $\begin{pmatrix}
            1 & 0.60 & 0.86 & 0.31 & 0.33 \\
             & 1 & 0.58 & 0.13 & 0.57 \\
             &  & 1 & 0.35 & 0.29 \\
             &  &  & 1 & 0.01 \\
             &  &  &  & 1
        \end{pmatrix}$ &  \\
    \hline
AICc & $-$69422 &  \\
conv iter & 41 &  \\
time (sec) & 91 &  \\
    \hline
\end{tabular}
\label{msVG results}
\end{table}

The model is then extended to include an AR(1) term in the mean to describe the autocorrelation of the five return series.
Similarly, results for the estimated parameters and their standard errors are given in Table \ref{msVG AR1 results}:

\begin{table}[H]
\caption{Results for the Multivariate Skewed VG model with AR(1) term}
\begin{tabular}{|l|c|c|}
    \hline
 & estimate & standard error \\
    \hline \hline
$\bm\beta_0^T$ &
        $10^{-4}\begin{pmatrix}
            16 & 13 & 10 & 14 & 2
        \end{pmatrix}$ &
        $10^{-4}\begin{pmatrix}
            3 & 3 & 3 & 2 & 6
        \end{pmatrix}$ \\
    \hline
$\bm\beta_1$ &
        $10^{-2}\begin{pmatrix}
            -8 & 35 & -18 & -1 & 2 \\
            8 & -11 & -4 & -7 & -1 \\
            -14 & 37 & -13 & -1 & -0.4 \\
            -4 & 43 & 24 & -21 & -2 \\
            10 & -27 & 1 & -9 & -5
        \end{pmatrix}$ &
        $10^{-2}\begin{pmatrix}
            4 & 3 & 5 & 3 & 1 \\
            3 & 3 & 4 & 2 & 1 \\
            3 & 3 & 3 & 2 & 1 \\
            2 & 2 & 3 & 2 & 1 \\
            6 & 6 & 8 & 5 & 2
        \end{pmatrix}$
         \\
    \hline
$\bm\Sigma$ &
        $10^{-5}\begin{pmatrix}
            17 & 10 & 12 & 3 & 13 \\
             & 14 & 8 & 2 & 17 \\
             &  & 12 & 3 & 10 \\
             &  &  & 7 & 2 \\
             &  &  &  & 64
        \end{pmatrix}$ &
        $10^{-6}\begin{pmatrix}
            7 & 5 & 5 & 3 & 9 \\
             & 5 & 4 & 2 & 8 \\
             &  & 5 & 2 & 7 \\
             &  &  & 3 & 5 \\
             &  &  &  & 25
        \end{pmatrix}$
         \\
    \hline
$\bm\gamma^T$ &
        $10^{-4}\begin{pmatrix}
           -13 & -12 & -8 & -12 & -0.2
        \end{pmatrix}$ &
        $10^{-4}\begin{pmatrix}
            4 & 4 & 4 & 3 & 8
        \end{pmatrix}$
         \\
    \hline
$\nu$ & 1.45 & 0.057 \\
    \hline \hline
Corr &
        $\begin{pmatrix}
            1 & 0.65 & 0.85 & 0.29 & 0.38 \\
             & 1 & 0.64 & 0.22 & 0.56 \\
             &  & 1 & 0.33 & 0.35 \\
             &  &  & 1 & 0.09 \\
             &  &  &  & 1
        \end{pmatrix}$ &  \\
    \hline
AICc & $-$70866 &  \\
conv iter & 50 &  \\
time (sec) & 111 &  \\
    \hline
\end{tabular}
\label{msVG AR1 results}
\end{table}

\begin{figure}[htbp]
  \begin{center}
    \subfigure[DAX]{\label{}\includegraphics[width=0.32\linewidth]{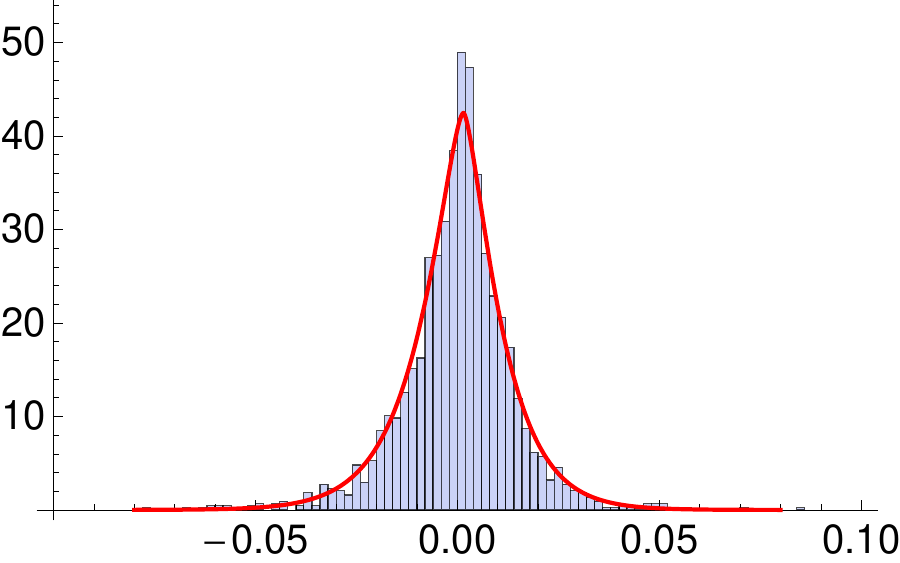}}
    \subfigure[S\&P 500]{\label{}\includegraphics[width=0.32\linewidth]{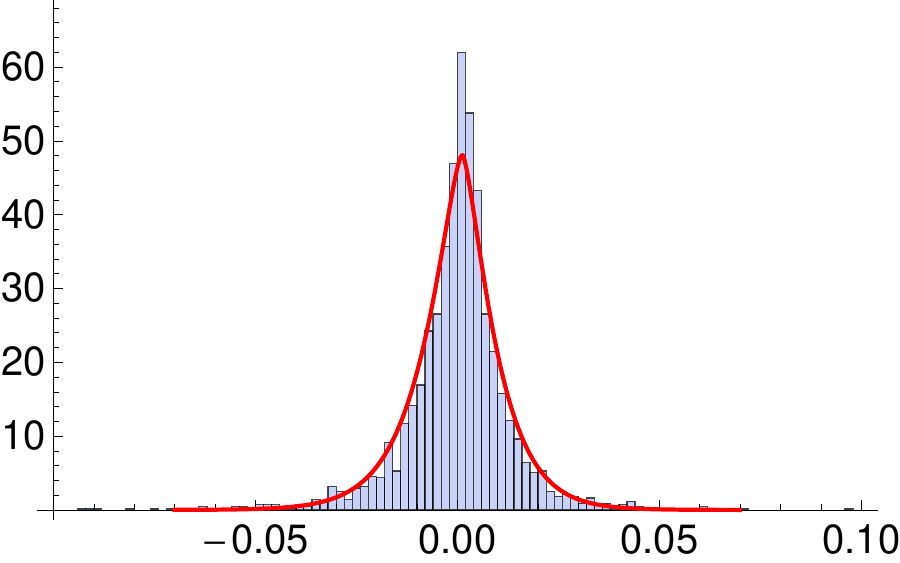}}
    \subfigure[FTSE 100]{\label{}\includegraphics[width=0.32\linewidth]{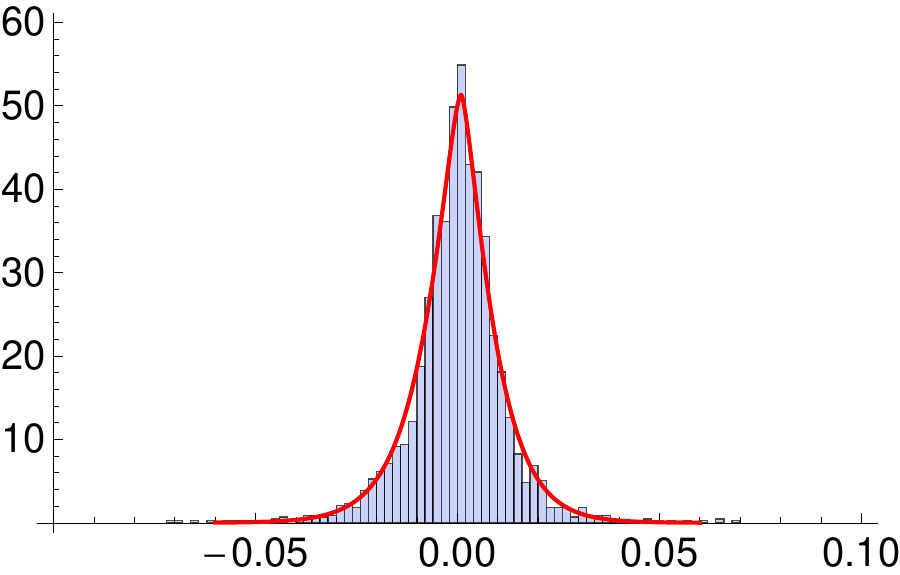}}
    \subfigure[AORD]{\label{}\includegraphics[width=0.32\linewidth]{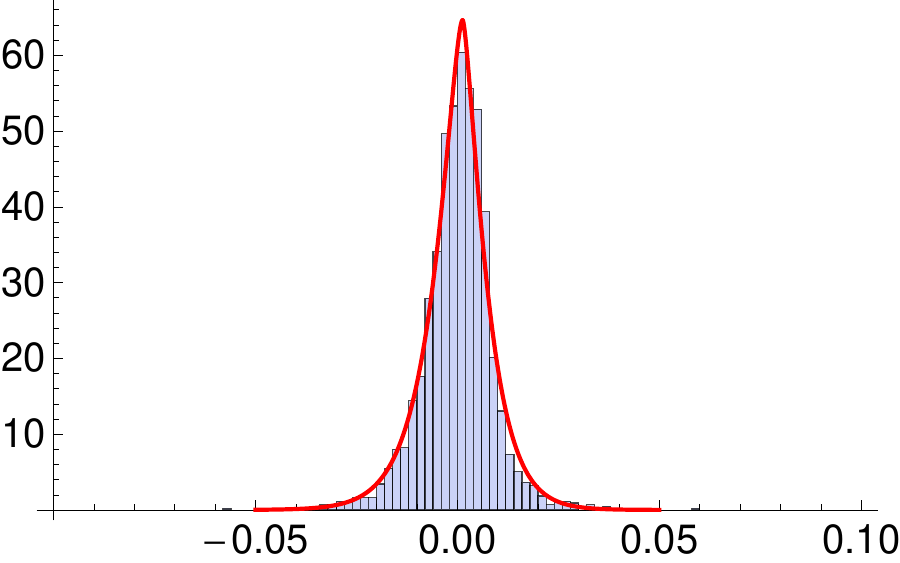}}
    \subfigure[CAC]{\label{}\includegraphics[width=0.32\linewidth]{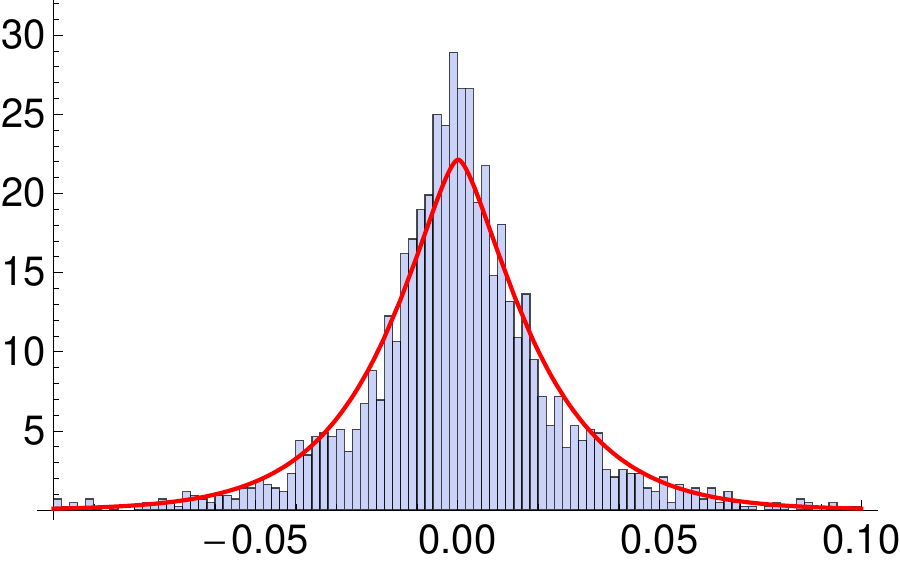}}
  \end{center}
  \caption{Histograms of AR(1) filtered residuals for DAX, S\&P 500, FTSE 100, AORD, CAC  data sets}
  \label{msVG AR(1) plots}
\end{figure}

The two sets of model parameters are not directly comparable.
However the estimates of $\bm \Sigma$ and correlation matrices are very similar while $\bm\gamma$ are some what similar.
The standard error for the estimate $\bm\beta_1$ in MSVG AR model suggests that only the second column has all parameters being significant.
Since the second column of the $\bm\beta_1$ matrix has relatively larger values, all five stocks are strongly lag-one cross-correlated with S\&P 500.
It is not surprising to know that the returns in S\&P 500 has the most impact on each of the five returns the next day because S\&P 500 has been shown to be a strong predictor for a number of market indices.
This is due to its large market share and its minimal real time difference with lag-one return of other markets.
Comparing the AICc of the two models, the MSVG AR model provides better model fit
which is further illustrated in the histogram of residuals in Figure \ref{msVG AR(1) plots}
after filtering out the AR(1) term. Fitted marginal pdfs with $\bm\beta_0$ as the mean of the MSVG model are added to the figure to facilitate comparison.
Overall the two MSVG models provide good fits to the data despite occasionally, the peaks of the histogram and fitted pdf around the means does not match exactly possibly due to the rather strong assumption of a common shape parameter $\nu$ across all components. This is a limitation of the MSVG distribution.

\begin{figure}[htbp]
  \begin{center}
    \subfigure{\label{}\includegraphics[width=0.6\textwidth]{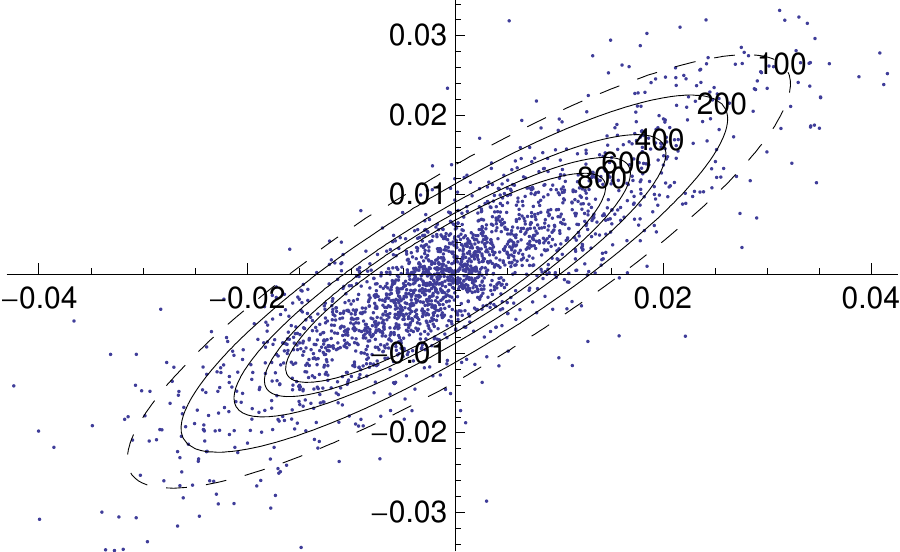}}
  \end{center}
  \caption{Fitted contour plot of AR(1) filtered DAX and FTSE 100 data sets}
  \label{contourplot DAX and FTSE with AR(1) filter}
\end{figure}

From the correlation matrices of the two models, the pair of market indices DAX and FTSE 100 are highly correlation ($\rho_{13}\simeq 0.85$), possibly due to the strong competitiveness of the German and UK markets as they are the major stock markets in Europe.
This gives rise to similar cross-correlation for both DAX and FTSE 100 with the other stocks.
The model fit of the MSVG AR model for DAX and FTSE 100 bivariate return data is displayed in the contour plot in Figure \ref{contourplot DAX and FTSE with AR(1) filter}.
Since the parameter estimates are quite similar to the five dimensional empirical study, we omit reporting these results.
Overall the model captures the strong correlation between the two stocks, giving a good overall fit to the DAX and FTSE 100 data set.

\section{Conclusion}     \label{Section: Conclusion/Discussion}

The MCECM and ECME algorithms have been proposed to obtain the ML estimates of multivariate Student-t distribution \citep{Liu1995}. We advance the algorithms to HECM algorithm and demonstrate its applicability to the MSVG model with AR mean function. Our proposed HECM algorithm with two adjustments solves three technical issues in application. \par \vspace{3mm}

Firstly, HECM algorithm improves the overall computational efficiency by combining the speed of MCECM algorithm at the start of iterations with the stability of ECME algorithm at latter iterations. Secondly, when the shape parameter is small and observations cluster closely to the mean, the problem of unbounded MSVG density leading to diverged estimates $\widehat{1/\lambda_i}$ and $\widehat{\log \lambda_i}$ can be resolved by capping the density within certain range of $\Delta$. We study
the optimal choices of $\Delta$ ranges for dimension $d=2,3$. More studies are needed to verify the choices of $\Delta$ for higher dimensions.
Lastly, we add an extra E-step after updating $(\bm\mu^{(t+1)}, \bm\gamma^{(t+1)})$ but before updating $\bm\Sigma^{(t+1)}$ to improve the stability of the $\hat{\bm\Sigma}$ estimate when the MSVG density is again unbounded. \par \vspace{3mm}

The HECM algorithm is then applied to fit the MSVG model with an AR(1) mean structure to describe the dynamics in financial time series.
To improve the model flexibility and predictability, the HECM algorithm can be easily extended to models with AR($p$) terms.
Moreover, the algorithm can be further enhanced to popular financial time series models such as GARCH models \citep{Bollerslev1986} and with leverage effect \citep{Engle1993}.
Some distribution families such as the multivariate skew Student-t distribution or multivariate skew generalised hyperbolic distribution which nests the MSVG distribution may be considered because they can be expressed in normal mean-variance mixtures representation.
While these extensions improve the applicability of the algorithm, it is very challenging as there is no close-form solution for parameters in the mean function and hence more layers of iterations are required.  \par \vspace{3mm}

In summary, we show that the HECM algorithm improves the computation efficiency in the ML estimation for the complicated MSVG distribution in normal mean-variance mixtures representation. However, we also remark the limitation of the MSVG model with one shape parameter for all components. In practice, different time series may follow distributions with different shape parameters. Hence, to improve the model applicability, it is necessary to allow one shape parameter for each component resulting in a modified MSVG model, similar to the GARCH models in \citet{Choy2012}. \par \vspace{3mm}

\section*{Acknowledgments}
I would like to thank John Ormerod for his helpful comments which leads to improvement of the paper.


\end{document}